\documentstyle[procl,epsfig]{article}

\def\xbj{x_{\rm Bj}}
\def\updn{\uparrow\downarrow}
\def\half{{\textstyle{1\over 2}}}
\def\ycut{y_{\rm cut}}
\def\beqn{\begin{eqnarray}}
\def\eeqn{\end{eqnarray}}
\def\beq{\begin{equation}}
\def\eeq{\end{equation}}
\def\cF{{\cal F}}
\def\cR{{\cal R}}
\def\cP{{\cal P}}
\def\cG{{\cal G}}
\def\oQ{\overline{Q}}
\def\as{\alpha_s}
\def\msb{\overline{\rm MS}}
\def\asb{\bar{\alpha}_s}

\def\GeV{{\rm GeV}}

\def\TeV{{\rm TeV}}
\def\lapproxeq{\lower .7ex\hbox{$\;\stackrel{\textstyle <}{\sim}\;$}}
\def\gapproxeq{\lower .7ex\hbox{$\;\stackrel{\textstyle >}{\sim}\;$}}

\begin{document}

\title{SUMMARY}

\author{W.J. STIRLING}

\address{Departments of Mathematical Sciences and Physics, 
University of Durham,\\ 
South Road, Durham DH1~3LE, UK}

\maketitle\abstracts{Some of the new experimental 
results and theoretical developments
presented at the Workshop on 
{\it Deep Inelastic Scattering and Related Phenomena} are reviewed.}

\section{Introduction}
\label{sec:intro}
The present series of Workshops on deep inelastic scattering and related
high-energy processes began in Durham in 1993. At that time results from the
HERA $ep$ collider were beginning to appear, and a forum where experimentalists
and theorists could get together to discuss these and other related measurements
seemed appropriate. The meeting was a resounding success. The quality and 
quantity of the physics, together with the enthusiasm of the participants, 
all pointed towards the establishment of a `deep inelastic scattering'
workshop as an annual event. The Eilat (1994) and Paris (1995) meetings
confirmed the Workshop as a truly international meeting, 
and one of the most important
in the high energy physics calendar.
 This tradition has been continued in Rome, with a record
 number of participants and a wealth of interesting physics.
 
Although HERA provided the original motivation, one of the keys to the success
of the DIS Workshops is the way they draw together the whole deep inelastic
scattering community, with fixed-target experiments playing an equally important
role. The hadron collider community has also been well represented, 
illustrating the complementarity of lepton-hadron and hadron-hadron collisions
in providing information on hadron structure.

There have been four main strands to the physics discussed at this Workshop:
(i) investigating the parton structure of the proton and photon as revealed in 
high-energy lepton-hadron and photon-hadron collisions respectively; (ii) 
understanding the origin  of those events which are both deeply inelastic
and diffractive; (iii) studying detailed QCD dynamics by means of 
particular hadronic final states (jets, heavy flavours, \ldots); and (iv)
unravelling the spin structure of the  nucleon by means of polarized
deep inelastic scattering experiments. In this brief review I will attempt
to highlight some of the new results in these different areas, together
with their theoretical implications. The choice is necessarily restricted
(lack of space and personal expertise being the main constraints)
and does not come close to doing justice to all the interesting physics
which has been presented and discussed. Nevertheless, I hope it will give
 a flavour of what is without question one of the most
 important frontiers of particle physics today.

\section{Proton structure}
\label{sec:structure}
\begin{figure}[htb]
\begin{center}
\mbox{\epsfig{figure=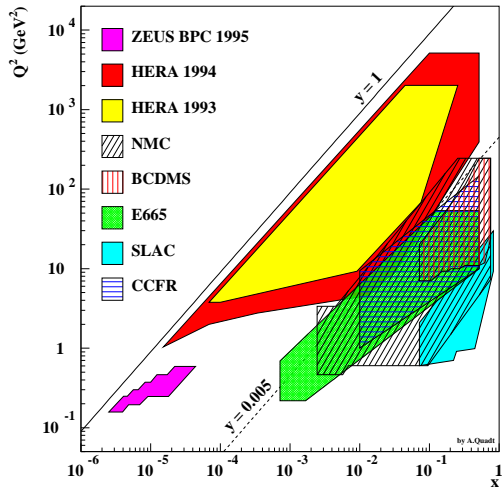,height=7cm}}
\caption{$x,Q^2$ phase space covered by various recent DIS experiments.}
\label{phase}
\end{center}
\end{figure}

The traditional method of obtaining information on the parton structure
of the nucleon is through measurements of deep inelastic structure functions.
The key variables (for example, for the process ${\rm e}^-(k) + 
{\rm p}(p) \to {\rm e}^-(k'=k-q) + X$) are $x$ and $Q^2$, where
\beq
Q^2 = - q^2 =  -(k-k')^2\; , \qquad x = Q^2/2p\cdot q \; .
\label{eq:xqdef}
\eeq
Structure functions $\cF_i(x,Q^2)$ are then obtained from the differential
scattering cross section, $d^2\sigma/dxdQ^2$. The main advance in recent years
has been the dramatic increase in the range of $x$ and $Q^2$ covered by 
experiment. With improvements in luminosity and detectors,
HERA has been able to measure $F_2^{ep}$ down to $x \sim 10^{-5}$. At the same time,
the $Q^2$ range has been extended  both at the upper {\it and} lower ends.
For the former, this leads to an increase in quark substructure limits
(see Section~\ref{subsec:hiq}) and tests of perturbative QCD (pQCD)
evolution.
Data at low $Q^2$ are important  for providing a bridge to the 
fixed-target data, and also for understanding the 
perturbative/nonperturbative transition. The region covered by the most recent
HERA and fixed-target data is illustrated in Fig.~\ref{phase}.
\subsection{New structure function measurements}
\label{subsec:newdata}
Two significant improvements have been reported at this Workshop
\cite{h1sum,zeussum}. The 1994 HERA $F_2$ data now overlap with the fixed-target
data, in the region $Q^2 \sim 10 - 100\;\GeV^2$, $x \sim 0.01 - 0.1$,
and the total ranges now spanned are
\beq
1.5 \lapproxeq Q^2 \lapproxeq 5000 \; \GeV^2, \qquad 3\times 10^{-5}
\lapproxeq   x \lapproxeq 0.3\; .
\eeq
As the kinematic region of the HERA $F_2$ data continues to grow, two
notable features persist:
\begin{itemize}
\item $F_2$ rises at small $x$ for all $Q^2$;
\item NLO DGLAP evolution provides an excellent description of the $Q^2$
dependence (see next Section).
\end{itemize}

\begin{figure}[htb]
\begin{center}
\mbox{\epsfig{figure=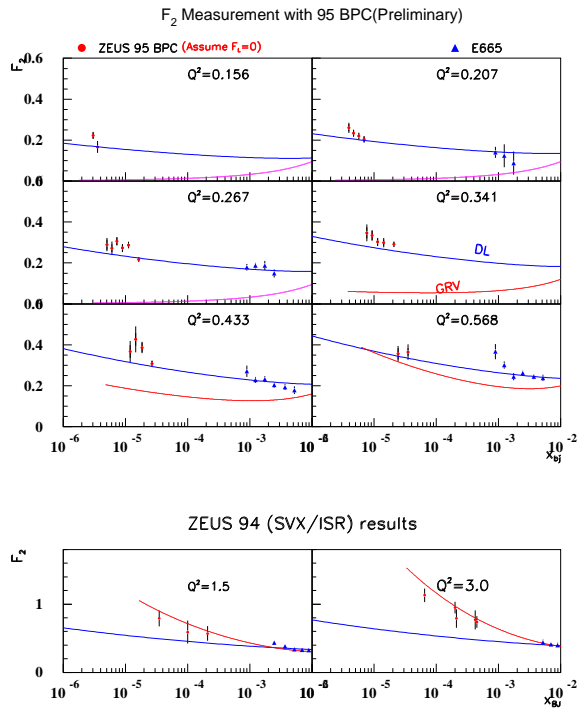,height=10cm}}
\caption{Preliminary ZEUS 1995 BPC $F_2$ measurements (top), together
with ZEUS 1994 SVX/ISR data at higher $Q^2$ (bottom), from
Ref.~\protect\cite{zeusbpc}. The predictions of the theoretical models
from Refs.~\protect\cite{dola,grv94} are also shown.}
\label{fig:zeusbpc}
\end{center}
\end{figure}
\begin{figure}[htb]
\begin{center}
\mbox{\epsfig{figure=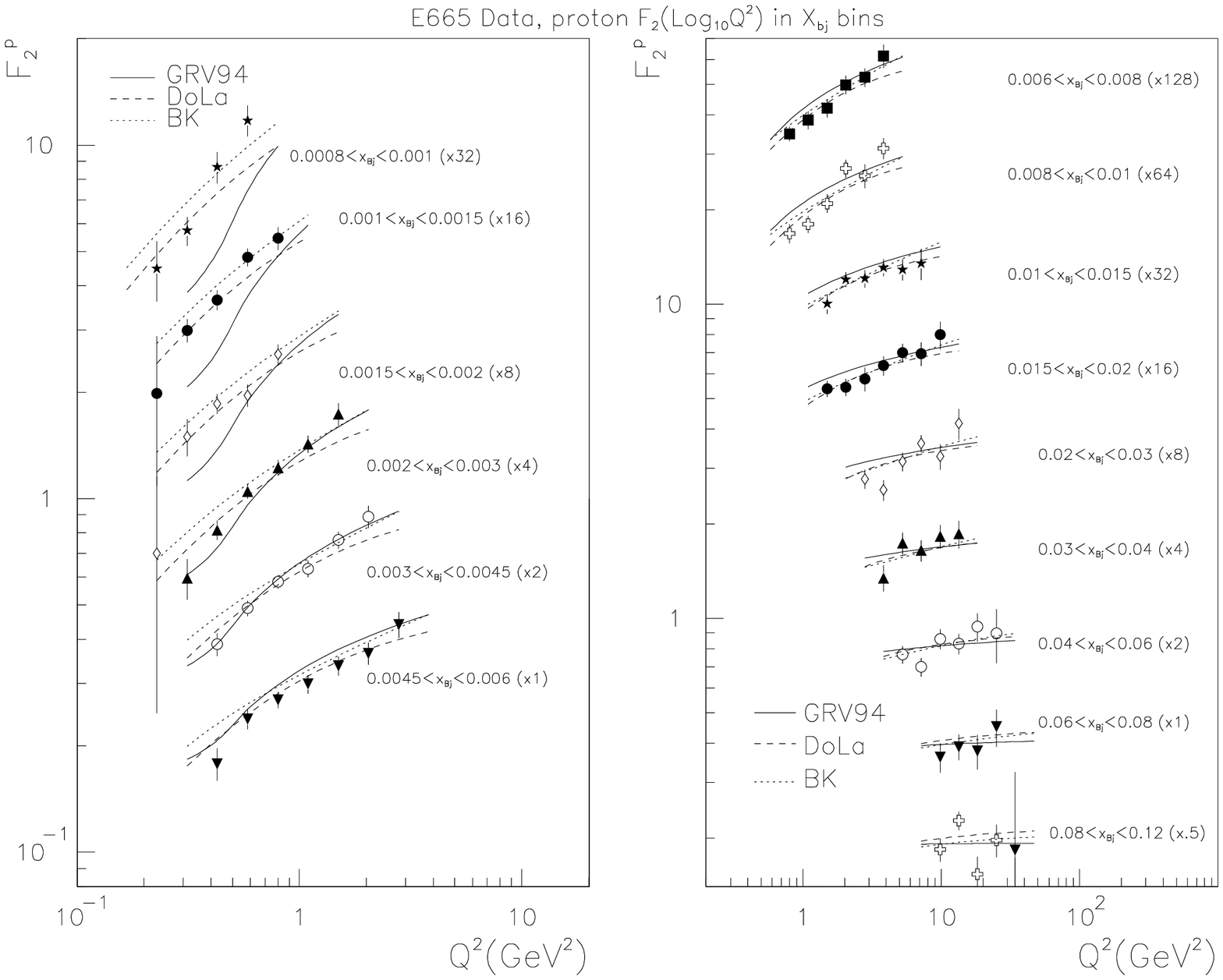,height=8cm}}
\caption{$F_2^p$ measurements by the E665 collaboration
\protect\cite{e665}, with various theoretical model predictions
\protect\cite{dola,grv94,bk}.}
\label{fig:e665}
\end{center}
\end{figure}

The kinematic range of the ZEUS data has been further extended by the
installation in 1995 of a beam pipe calorimeter. This allows electron
and positron
scattering at smaller angles to be measured, which in turn leads, via
Eq.~(\ref{eq:xqdef}), to smaller $Q^2$ values. Preliminary
$F_2$ measurements, reported at the Workshop~\cite{zeusbpc},
are shown in Fig.~\ref{fig:zeusbpc}. The corresponding $x,Q^2$ range is
\beq
\mbox{ZEUS BPC}: \qquad  0.16 \lapproxeq Q^2 \lapproxeq 0.57 \; \GeV^2,
\qquad    x \sim 10^{-5}\; .
\eeq
Fig.~\ref{fig:zeusbpc} also illustrates the complementarity of the HERA and
fixed-target structure function measurements. At the same low $Q^2$
values the E665 collaboration~\cite{e665} have measured $F_2^{\mu p}$ 
for $x \sim 10^{-3} - 10^{-2}$, see Fig.~\ref{fig:e665}.
All these data are important for constraining models of structure
functions at low $Q^2$.  For example, the Donnachie-Landshoff
Regge-based model~\cite{dola} appears to give a good description
of the $Q^2 \lapproxeq 1\; \GeV^2$ data, interpolating between the
ZEUS and E665 data with a slowly rising $F_2 \sim x^{-0.08}$ form.
However this model does {\it not} include pQCD $Q^2$ evolution, and
therefore fails to describe the steepening of $F_2$ with increasing
$Q^2$ which is apparent in the SVX/ISR data in Fig.~\ref{fig:zeusbpc}.
A comprehensive and critical review of models for $F_2$ at low $Q^2$
can be found in Ref.~\cite{levy}
Other new structure function data reported at the Workshop
include updated measurements of $F_2^{p,d}$ and $F_2^n/F_2^p$
from the NMC collaboration~\cite{nmc}. These are particularly
useful for constraining the medium-$x$ quark distributions (in
particular the $u/d$ ratio) and will be incorporated in forthcoming
`global fit' analyses.

\subsection{DGLAP evolution}
\label{subsec:nlodglap}
One of the central tenets of perturbative QCD is that the $Q^2$ evolution
of structure functions $\cF_i(x,Q^2)$ is determined by the DGLAP
equations, provided that $Q^2$ is sufficiently large such that higher-twist
($\propto 1/Q^2$) contributions can be neglected. 
More precisely, the theory predicts the factorization scale
dependence of the quark and gluon distributions,
$q(x,\mu^2)$ and $g(x,\mu^2)$,
\beq \label{dglap}
\mu^2{\partial \over \partial \mu^2}\;
 \left(\begin{array}{c} q \\ g \end{array}\right)
\; = \; {\as(\mu^2) \over 2 \pi}\;  \left(\begin{array}{cc}
 P_{qq} & P_{qg}  \\
 P_{gq} & P_{gg}
\end{array}\right) \otimes
\left(\begin{array}{c}q \\ g\end{array}\right) \; ,
\eeq
where $\otimes$ denotes a convolution integral and
the splitting functions have a perturbative expansion
\beq 
P_{ab}(x,\as) = \as P_{ab}^{(0)}(x) + \as^2 P_{ab}^{(1)}(x) + \ldots\; .
\label{splitho}
\eeq
The structure functions are obtained as linear combinations
of the parton distributions:
\beqn
\cF_i(x,Q^2) &= &\int_x^1dy/y \left[ \sum_q  q(y,Q^2) \otimes
C_{q,i}(x/y,\as(Q^2)) \right. \nonumber \\
& & \left. + g(y,Q^2) \otimes C_{g,i}(x/y,\as(Q^2))\right]\; .
\eeqn
For example, for $\cF_i = x^{-1}F_2^{ep}$ the coefficient functions are
$C_{q,i}(z,\as) = e_q^2\delta(1-z) + O(\as)$, $C_{g,i}(z,\as) = O(\as)$.
\begin{figure}[htb]
\begin{center}
\mbox{\epsfig{figure=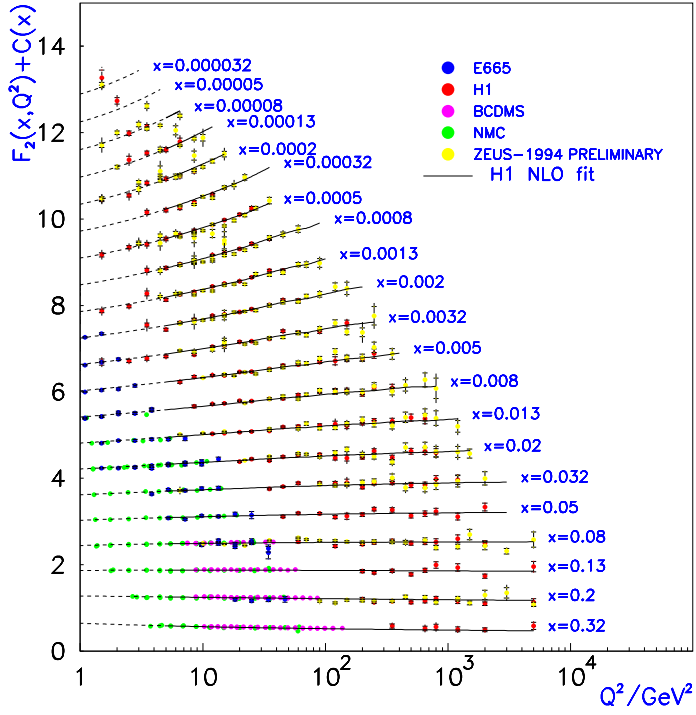,height=10cm}}
\caption{The $Q^2$ dependence of  $F_2$  at small and
medium $x$, together with a NLO QCD fit,
from the H1 collaboration \protect\cite{h1sum}.}
\label{fig:f2nlo}
\end{center}
\end{figure}
Once the distributions $f_i(x,Q^2)$ ($i=q,g$) and $\as$ are 
specified~\footnote{In
practice the value of $\as(M_Z^2)$ rather than $\as(Q_0^2)$ is 
used to quantify the
strong coupling.}  at some
`starting' scale $Q_0^2$, the theory predicts the distributions at any
$Q^2$ for which perturbation theory is valid.
 In practice the starting distributions are determined from the
data by means of a {\it global fit}, see for example
Refs.~\cite{dick,wuki}. It is remarkable that only
relatively simple functional forms are required, for example
\beq
x f_i(x,Q_0^2) = A_i x^{-\lambda_i} (1 + \epsilon_i \sqrt{x}
+ \gamma_i x) (1-x)^{\eta_i} \; .
\label{eq:starting}
\eeq
The splitting and coefficient functions are known exactly at leading and
next-to-leading order (see for example Ref.~\cite{book} for a compilation
and list of references). Truncating at this order defines the NLO DGLAP
system of equations.

Several examples of NLO DGLAP fits to structure function data have been
presented at the Workshop. Considering the wide range of processes and
energies, the quality of the description is excellent. An example from
the H1 collaboration~\cite{h1bassler} is shown in Fig.~\ref{fig:f2nlo}.
What information can be extracted from such fits? At large $x$
the DGLAP equation for $F_2$ reduces to $\partial F_2 / \partial \ln Q^2
\sim \as F_2 \otimes P_{qq}$, and a precision measurement of $\as$ can
be made, see Section~\ref{sec:alphas} below.
At small $x$ the gluon is the dominant parton and so
$\partial F_2 / \partial \ln Q^2 \sim \as g \otimes P_{qg}$.
Thus while $F_2$ directly measures the {\it quark} distribution
at small $x$, its $Q^2$ variation measures the {\it gluon}.

The shape of the starting distributions (Eq.~(\ref{eq:starting})) at small $x$
is determined by the parameters $\lambda_i$.  The dominant partons
here are the gluons and sea quarks, for which we may write
\beq 
xg(x,Q_0^2) \; \sim \; A_g  x^{-\lambda_g}\; , \qquad
xq_S(x,Q_0^2) \; \sim \; A_S  x^{-\lambda_S}\; .
\label{lipa}
\eeq 
It is interesting to see how  $\lambda_g$ 
and $\lambda_S$, obtained from fits to data,
 have evolved with time. Table~\ref{tevolution} shows these
parameters for various recent MRS analyses.~\footnote{Since all the fits
listed have been performed at NLO in the $\overline{\rm MS}$ scheme, the
parameters {\it corresponding to the same $Q_0^2$} are directly comparable.}
\begin{table}[htb]
\begin{center}
\begin{tabular}{|rcccc|}  \hline
\rule[-1.2ex]{0mm}{4ex} &  set & $Q_0^2$ &    $\lambda_S$  & $\lambda_g$  \\
\hline
\rule[-1.2ex]{0mm}{4ex}$<$1993  &        &     4  & 0       &    0       \\
\rule[-1.2ex]{0mm}{4ex}   1993  & D$_-'$ &     4  & $\half$ &  $\half$   \\
\rule[-1.2ex]{0mm}{4ex}   1994  & A      &     4  &    0.3  &  0.3       \\
\rule[-1.2ex]{0mm}{4ex}   1995  & A$'$   &     4  &  0.17   & 0.17       \\
\rule[-1.2ex]{0mm}{4ex}   1995  &  G     &     4  &  0.07   &   0.31      \\
\rule[-1.2ex]{0mm}{4ex}   1996 &  R$_1$ &     1  &   0.14      & -0.41   \\
   \hline
\end{tabular}
\caption{Evolution in time of the parameters determining the small-$x$
behaviour of the MRS quark and gluon distributions.\label{tevolution}}
\end{center}
\end{table}
Before HERA data became available, it was traditional to assume
Regge-motivated flat starting distributions ($\lambda \approx 0$). Subsequently,
theoretical studies of the BFKL equation for resumming large logarithms
of $1/x$ (see Section~\ref{subsec:bfkl}) suggested that the behaviour could 
in fact be much steeper
($\lambda \simeq 12\as\ln2/\pi \approx 0.5$). The early HERA data showed
a somewhat less steep rise at small $x$, with fits giving $\lambda \approx 0.3$.
More recent HERA data show a preference for {\it different}
$\lambda_S$ and $\lambda_g$ values. In fact in the most recent MRS fits
\cite{dick},
where the minimum $Q^2$ of the fitted data is extended down to $1.5\;
\GeV^2$,  the gluon is `valence-like' at the starting scale $Q_0^2 = 1\;
\GeV^2$. Even more significant is the fact that the slope of the sea quark
distribution is close to the Regge prediction obtained from
fits to the energy dependence of total hadronic cross sections~\cite{dola}:
\beq
xq_S \; \sim \;   x^{1- \alpha_{\cal P}(0) \approx -0.08}\
\mbox{at} \ Q_0^2 =  O(1\; \GeV^2)\; .
\label{regge}
\eeq
However the agreement between the fitted value of $\lambda_S$ and
$\alpha_{\cal P}(0)-1$ should not be taken too literally. The former
is an unphysical parameter, depending to some extent
on the factorization scheme,
the value chosen for $Q_0^2$, and the parametric form chosen for the
starting distributions. Another problem with the Regge interpretation
is that one would naively expect the gluon distribution to exhibit
the same asymptotic behaviour for $x \to 0$. One can impose by hand the
constraint $\lambda_S=\lambda_g$, in which case 
the common value $0.04$ (at $Q_0^2 = 1\; \GeV^2$) is obtained~\cite{dick}, 
but the quality of the fit is worse than when 
the parameters are allowed to be different.

\begin{figure}[htb]
\begin{center}
\mbox{\epsfig{figure=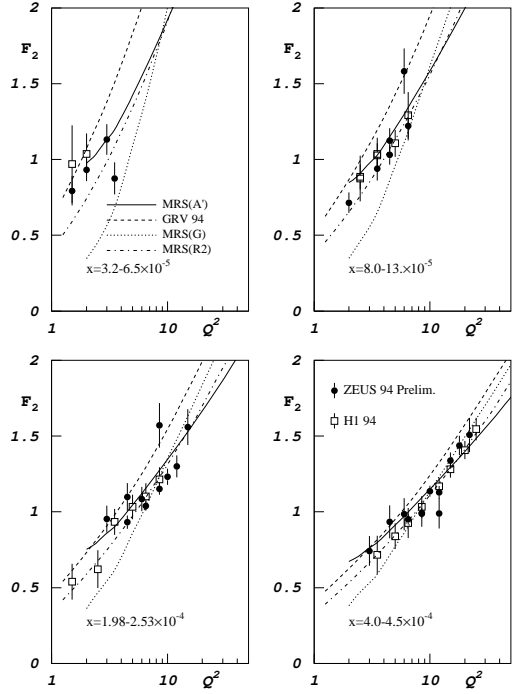,height=10cm}}
\caption{The description of the most recent HERA $F_2^{ep}$ data
by the  `old' parton distributions of Refs.~\protect\cite{grv94,mrsg},
and the new MRS(R$_2$) fit \protect\cite{dick}.}
\label{fig:newold}
\end{center}
\end{figure}

As Table~\ref{tevolution} implies, the continual improvement in the structure
function data requires a concomitant updating of the parton distributions
derived from global fits. Thus `new' data is able to discriminate
between `old' sets of partons. This is illustrated in Fig.~\ref{fig:newold},
which shows recent HERA $F_2$ data compared with the 1995 MRS
\cite{mrsg} and 1994 GRV~\cite{grv94}  `dynamical parton' predictions.
Also shown is the new MRS(R$_2$) fit~\cite{dick} which includes these
data. Notice how the MRS(G) and GRV(94) predictions are now disfavoured -- they
predict too strong a $Q^2$ dependence at small $x$. It will be
interesting to see if the agreement between the dynamical parton model
and the data can be restored by adjusting the parameters of the model,
for example the value of the starting scale $\mu_0^2$ at which the
distributions assume a valence-like form.

\begin{figure}[htb]
\begin{center}
\mbox{\epsfig{figure=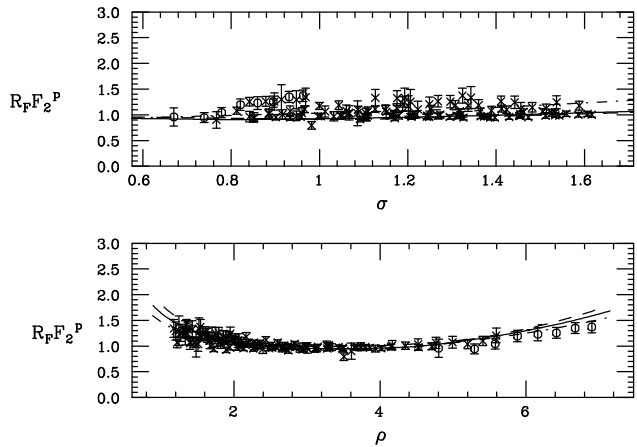,height=6cm}}
\caption{H1 $F_2$ data compared with the `double asymptotic scaling'
prediction \protect\cite{forte}. The curves are the full NLO best fit
and  have $\rho = 2,3,4$ in the 
$\sigma$ plot, and $\sigma = 1, 1.2, 1.4$ in the $\rho$ plot.}
\label{bf2}
\end{center}
\end{figure}

We can summarize the results of this section by the statement
that the small-$x$ structure function data are consistent with DGLAP
evolution starting from a soft input for both quarks and gluons
at a  scale of order $1\; \GeV^2$. The fact that the input is soft
means that we can obtain a simple analytic approximation to the
solutions of the evolution equation in the (double) asymptotic limit 
$Q^2,\; 1/x \to \infty$~\cite{ZEE}. In its simplest form, 
this result follows from the
leading $x\to 0$  behaviour of the lowest-order splitting function matrix:
\beq \label{dglapsmx}
\mu^2{\partial \over \partial \mu^2}\;
 \left(\begin{array}{c} q \\ g \end{array}\right)
\; \simeq \; {\as(\mu^2) \over 2 \pi}\; \left(\begin{array}{cc}
 0 & 0  \\
{2C_F / x}  & {2C_A / x} 
\end{array}\right) \otimes
\left(\begin{array}{c}q \\ g  \end{array}\right) \; ,
\eeq
which leads, for sufficiently soft starting distributions, to the
prediction~\cite{ZEE} 
\beq 
F_2 \sim x \sum q \sim  \exp\left[       2 \sqrt{{C_A \as\over \pi}
\ln{Q^2\over Q_0^2} \ln{x_0\over x}   } \right] \; .
\label{eq:das}
\eeq
Ball and Forte have extended this `double asymptotic scaling'  result to include
subleading corrections and NLO contributions~\cite{BF,forte}. 
An example of a comparison between
 theory and experiment is shown in Fig.~\ref{bf2}. Here $F_2$ has been rescaled
 by a factor which is essentially the right-hand side of (\ref{eq:das}),
 computed at NLO, in 
 order to remove the leading asymptotic behaviour.  The variables 
 $\sigma,\rho$ are defined in terms of $x$ and $Q^2$ by 
\beq 
\sigma  =    \left[{  \ln(x_0/x)    \ln(t/t_0)  } \right]^{1/2}\; , \quad
\rho  =    \left[{  \ln(x_0/x)  \over  \ln(t/t_0)  } \right]^{1/2}\; ,
\eeq
where $t=\ln(Q^2/\Lambda^2)$.
The fact that the data lie approximately on universal, horizontal bands
 is a demonstration
of the validity of double asymptotic scaling. A slight breaking of the scaling
behaviour is evident at large and small $\rho$ in the lower 
plot, where, respectively, $Q^2$ becomes
too small and $x$ becomes too large for the approximation to be valid.
Here, however, the full NLO DGLAP prediction (indicated by the curves)
gives a very good description.

\subsection{Beyond NLO DGLAP at small $x$}
\label{subsec:beyondnlo}
Why does NLO DGLAP evolution work so well? To attempt to answer this,
we consider two types of correction which could in principle be important
in particular regions of $x,Q^2$ space. First, we note that 
the  statement that `perturbative QCD
evolution describes the data down to $Q^2 \sim 1\; \GeV^2$' must 
be qualified by the condition `at small $x$'. At large $x$ it has been 
clearly shown that higher-twist (i.e. $1/Q^2$) contributions to the
structure functions are important. In Ref.~\cite{ALAIN}, a combined
leading-twist/higher-twist analysis of structure function data, using 
the empirical form
\beq
F_2^{(HT)}(x,Q^2) = F_2^{(LT)}(x,Q^2)\ { C(x)\over  Q^2}\; ,
\label{ht}
\eeq
gave $C \approx 0.3\; \GeV^2/(1-x)$ at large $x$ and $C \approx 0$ at small
$x$.
The factor $(1-x)^{-1}$ is not unexpected, since
there are $j$ times more operators which contribute to the $j$th
moment of $F_2^{(HT)}$ than to $F_2^{(LT)}$. Therefore at large $j$ the moments
are expected to differ by a factor of $j$, corresponding to a factor  
$(1-x)^{-1}$ difference in the  contributions to the structure function.
Although the analysis of Ref.~\cite{ALAIN}
 was restricted to fixed-target data with $x \gapproxeq 0.1$
it would appear that the 
small-$x$ HERA structure functions with $Q^2 \gapproxeq 1.5\; \GeV^2$
are also relatively free from higher-twist contributions. In fact 
a global fit to the HERA ($Q^2 > 1.5\; \GeV^2$) $F_2$ data which includes  
a phenomenological higher-twist contribution of the form $(a^2/Q^2)F_2^{(LT)}$
gives a value for $a^2$ consistent with zero~\cite{dickpriv}.

\begin{figure}[htb]
\begin{center}
\mbox{\epsfig{figure=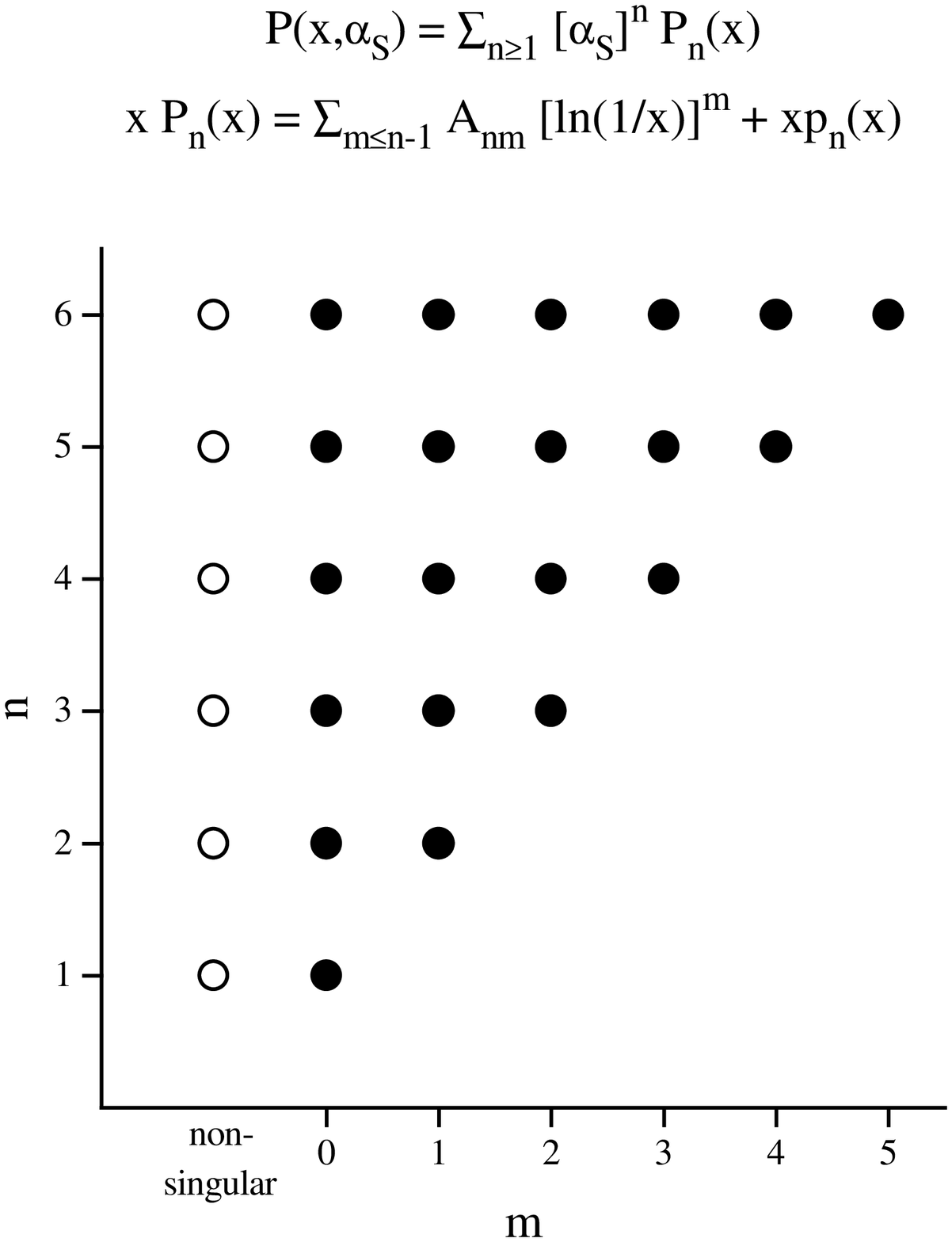,height=9cm}}
\caption{Pattern of $\ln(1/x)$ coefficients in the splitting functions
at small $x$.}
\label{fig:bfklcoeff}
\end{center}
\end{figure}
A second important correction to NLO DLGAP evolution at small $x$
comes from higher-order terms on the right-hand side of (\ref{splitho}).
As $x\to 0$, large $L_x \equiv \ln(1/x)$ logarithms appear in the splitting
functions and spoil the convergence. The theory and phenomenology of
these logarithms have been extensively discussed at this Workshop
\cite{many}. In general one can show that there are at most
$n-1$ large logarithms in the $n$th order splitting function, i.e. 
\beq
xP^{(n)}(x)  =  A_{n,n-1} L_x^{n-1} + A_{n,n-2} L_x^{n-2} + \ldots 
  + A_{n,0}  + xp^{(n)}(x) \; ,
\eeq
where $ p^{(n)}  $ is finite in the limit $x \to 0$.
The full splitting function is then
\begin{equation}
\label{expansion}
xP( x, \as ) = \sum_{n=1}^{\infty} \left( {\as \over 2 \pi} \right)^n
   \left[ \sum_{m=0}^{n-1} A_{n,m}  L_x^m + xp^{(n)}(x) \right]\; .
\end{equation}
This expression holds for all four splitting functions, $P_{gg}$, $P_{qg}$,
$P_{gq}$ and $P_{qq}$. The pattern of the coefficients is illustrated
in Fig.~\ref{fig:bfklcoeff}. The horizontal rows represent the complete
splitting function at a given order in perturbation theory, for example
LO$(n=1$), NLO$(n=2)$, etc. The leading diagonal with $n=m+1$
represents the set of leading $L_x$ contributions, and so on.

\begin{figure}[htb]
\begin{center}
\mbox{\epsfig{figure=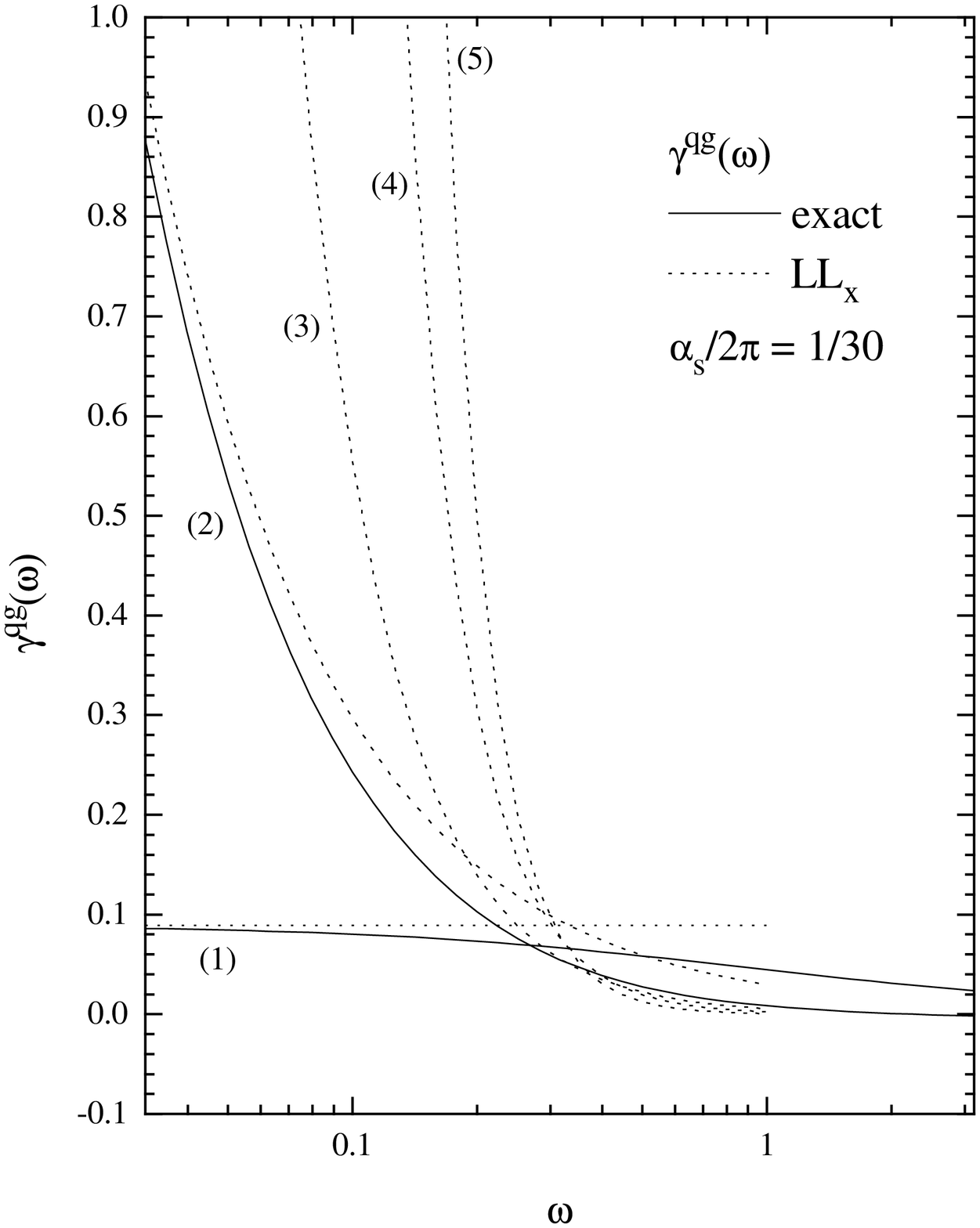,height=8cm}}
\caption{The anomalous dimension $\gamma^{qg}(\protect\omega)$, showing the
exact first- and second-order contributions, together with the 
leading-logarithm parts of the first five orders.}
\label{fig:anomqg}
\end{center}
\end{figure}

\begin{figure}[htb]
\begin{center}
\mbox{\epsfig{figure=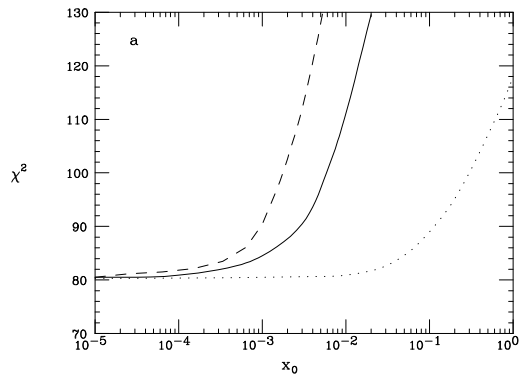,height=5cm}}
\caption{$\chi^2$ of various resummed $\ln(1/x)$ fits to the 
HERA $F_2$ data versus $x_0$, from Ref.~\protect\cite{forte}.}
\label{bf1}
\end{center}
\end{figure}

 Of particular phenomenological importance is the
$P_{qg}$ function, since at small $x$ $\partial F_2 / \partial\ln Q^2
\sim \as P^{qg}\otimes g$. To illustrate the importance of the
higher-order contributions, it is convenient to take moments,
$\gamma(\omega)= \int_0^1 dx\; x^{\omega}\; P(x)$, so that
$xP(x) \sim L_x^{n-1} \Rightarrow \gamma(\omega)\sim\Gamma(n)/\omega^n$.
The leading behaviour as $\omega \to 0$ (equivalently $x\to 0$) of
$\gamma^{qg}$ is found to be~\cite{CATHAU} (with $\asb = 3 \as/\pi$)
\beq
\gamma^{qg}(\omega) = n_f{\asb \over 9}   \left[ \left(1 + \ldots\right)
+ \asb \left(  {2.17 \over \omega}+ \ldots \right)
+ \asb^2 \left(  {2.30 \over \omega^2}+ \ldots \right) + \ldots \right] \; ,
\label{eq:expand}
\eeq
where at each order in perturbation theory subleading
terms down by one or more powers of $\omega$ have been suppressed. In fact beyond NLO
only the leading
$\asb(\asb/\omega)^n$ coefficients are known.
In the limit $\asb L_x \gg 1$ they resum to give the characteristic
BFKL `perturbative pomeron' behaviour~\cite{CATHAU}
\beq
x P_{qg} \sim x^{-4\ln 2\; \asb}\; .
\eeq
This asymptotic limit is, however, not relevant for the HERA structure function
data. The interesting question is how important in practice are the
contributions beyond NLO in (\ref{eq:expand}). Certainly the leading
logarithm part of these contributions appears to be large. This is
illustrated in Fig.~\ref{fig:anomqg}, which compares the exact leading-
and next-to-leading-order contributions with the leading-logarithm part
of the first five orders in perturbation theory.
It would appear from Fig.~\ref{fig:anomqg} that the higher-order terms
could be phenomenologically important. On the other hand, retaining only the
{\it leading} 
terms at each order (as in the phenomenological analyses performed so far)
could well overestimate their importance. This is evidently
true of the LO and NLO contributions in the region $0.1 \lapproxeq
\omega \lapproxeq 1$.
 Certainly,
there is no evidence that the data require such contributions.
One can quantify this statement~\cite{bfsum} 
by replacing $\ln(1/x)$ in the leading-logarithm
expansion by $\ln(x_0/x)\theta(x_0-x)$, and regarding $x_0$ as a parameter
to be determined by the data. Note that this replacement has the effect
of artificially introducing sub-leading logarithms, since $L_x^n \to L_x^n + n
\ln(x_0)L_x^{n-1} + \ldots\; $. Fig.~\ref{bf1} shows the $\chi^2$ of the fit to the
HERA $F_2$ data as a function of $x_0$ (in various resummation schemes, 
see~\cite{forte,bfsum}).
In each scheme the minimum value of $\chi^2$ is attained for $x_0 \to 0$,
indicating that the data prefer fits without the higher-order leading $\ln(1/x)$
contributions.

\subsection{`BFKL' description of $F_2$}
\label{subsec:bfkl}

In the double asymptotic limit discussed in Section~\ref{subsec:nlodglap},
the evolution equation  sums leading powers of 
$[\as\ln(1/x)\ln Q^2 ]^n$ generated by multigluon emission, and
the distributions  increase faster than any power of $\ln(1/x)$ 
as $x \rightarrow 0$, see (\ref{eq:das}). 
 The dominant region of phase space is where the gluons 
have strongly-ordered transverse momenta, $Q^2 \gg k^2_{Tn} \gg ... 
\gg k^2_{T1}$. However such evolution does not include {\it all} the leading terms 
in the small-$x$ limit.  It neglects those terms which contain the leading 
power of $\ln(1/x)$ but which are not accompanied by the leading power of 
$\ln Q^2 $.  The BFKL equation~\cite{BFKL}, on the other hand,
sums the leading $\ln(1/x)$ terms while retaining
 the full $Q^2$ dependence.  The integration is taken over 
the full $k_T$ phase space of the gluons, not just the strongly-ordered part.
  The result is most conveniently established for the gluon distribution
unintegrated over $k_T$, 
\beq
x g(x,Q^2) = \int^{Q^2} d k_T^2\;
\cG(x,k_T^2) ,
\eeq 
and is
\beq
\label{bfklresult} 
\cG(x,k_T^2) \;\; \sim \;\; h(k_T^2)x^{-\lambda} ,
\eeq 
 where $h \sim (k_T^2 )^{-1/2}$ at large $k_T^2 $ and
  $\lambda$ is the  the maximum eigenvalue of the kernel $K$  of the 
BFKL equation
\beq
\label{bfkleqn}
{\partial \cG \over\partial L_x } = \int d^2 k_T'\; K(\vec{k}_T,
\vec{k}_T')\; 
\cG(x,k_T^{\prime 2})  \; .
\eeq 
For fixed $\as$,  $\lambda = 12\ln 2 \as/\pi \approx 0.5$.
The prediction for $F_2$ is obtained by using the $k_T$-factorization
theorem~\cite{CCH}
\beq
F_2(x,Q^2)   =  \int_x^1 {d x' \over x'} \int {dk_T^2 \over k_T^4}
\cG\left({x\over x'},k_T^2\right) \overline{F}_2(x', k_T^2,Q^2) ,
\eeq
where $\overline{F}_2$ denotes the quark-box contribution
$\gamma g \to q \bar q$ for the scattering of a photon of virtuality
$Q^2$ off a gluon with longitudinal momentum fraction $x'$
and transverse momentum  $k_T$.

\begin{figure}[htb]
\begin{center}
\mbox{\epsfig{figure=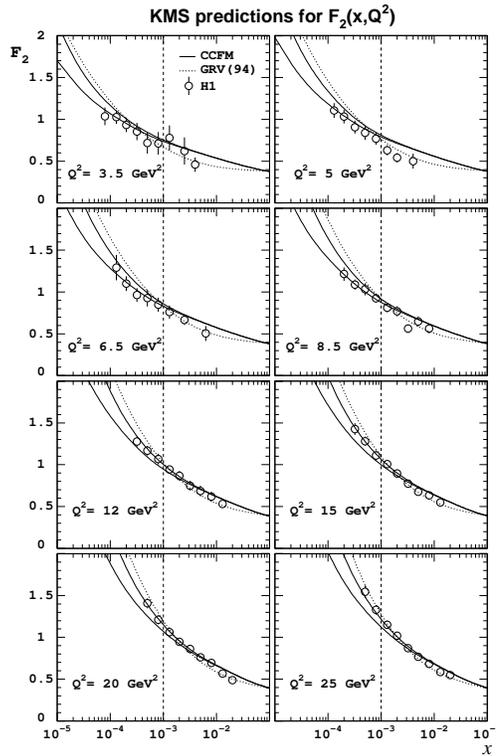,height=10cm}}
\caption{Constrained CCFM description of $F_2$ at small $x$ 
\protect\cite{kms}.}
\label{fig:kmsccfm}
\end{center}
\end{figure}
In recent years there have been several numerical analyses of the  
solution of the BFKL equation
and of the corresponding predictions for $F_2$, based on the 
above results (for a review see~\cite{alanbfkl}). The `naive' BFKL
prediction,
\beq
F_2(x,Q^2)   =  C(Q^2) x^{-\lambda} + F_2^{\rm NP} ,
\eeq
where $\lambda \approx 0.5$ and $F_2^{\rm NP}$ is a non-perturbative
background contribution which is constant at small $x$, would appear to 
give too steep a rise in $x$.
However there  has been significant recent progress, reported at this
Workshop~\cite{alanbfkl},  towards a `unified'
treatment which incorporates both DGLAP and BFKL dynamics, embodied in
the so-called CCFM equation, together with appropriate kinematic constraints.
A reasonable description of the HERA data is obtained~\cite{kms}, see
Fig.~\ref{fig:kmsccfm}. Here the continuous curves are the CCFM predictions
with and without the kinematic constraints.
From a theoretical point of view, the attraction of the BFKL approach
is that it attempts to explain the shape of the small-$x$ structure function
{\it from first principles}. However, at present one cannot discriminate
between the DGLAP and BFKL predictions on the basis of the $x$ and $Q^2$
dependence of $F_2$ alone. Furthermore, it is not clear how much
truly `BFKL' dynamics remains in the kinematically-constrained CCFM equation
applied to the HERA kinematic region. A key test would appear to be 
provided by more exclusive quantities, such as the average transverse 
hadronic energy 
in deep inelastic events (which is predicted to be larger in the BFKL
approach) and the cross section for producing forward, moderate $E_T$ jets
\cite{albert}.

\subsection{Gluon determinations}
\label{subsec:gluon}

Deep inelastic structure functions directly measure the quark 
distribution functions. The precision is set by the experimental
measurements, and is at the level of a few percent over a wide range in $x$,
at least for the $u$ and $d$ quarks. The gluon is much less well determined:
a variety of hard scattering processes provide measurements in particular
$x$ regions at different $Q^2$ scales, and momentum conversation pins down
the first moment to a few percent.

Several new results on the gluon determination have been reported at this 
Workshop. An overview is given in Ref.~\cite{rdam}. 
We  will concentrate here on those new results 
which are `theoretically precise', 
i.e. are performed using cross sections calculated to next-to-leading
order. This allows
the gluon to be defined and extracted according to a particular
factorization scheme (e.g. $\overline{\mbox{MS}}$), as for the quark
distributions.

\begin{figure}[htbp]
\begin{center}
\mbox{\epsfig{figure=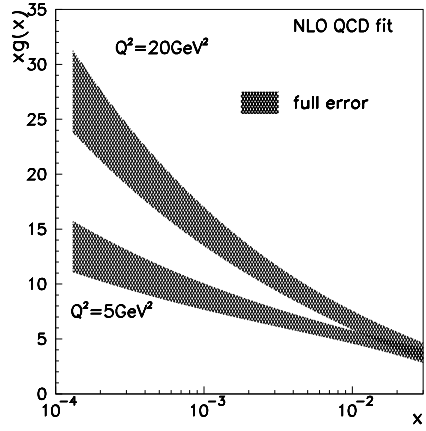,height=6cm}}
\caption{The gluon distribution obtained by H1 from a global NLO QCD fit
\protect\cite{h1nlofit}.}
\label{fig:h1gluon}
\end{center}
\end{figure}

\begin{figure}[htbp]
\begin{center}
\mbox{\epsfig{figure=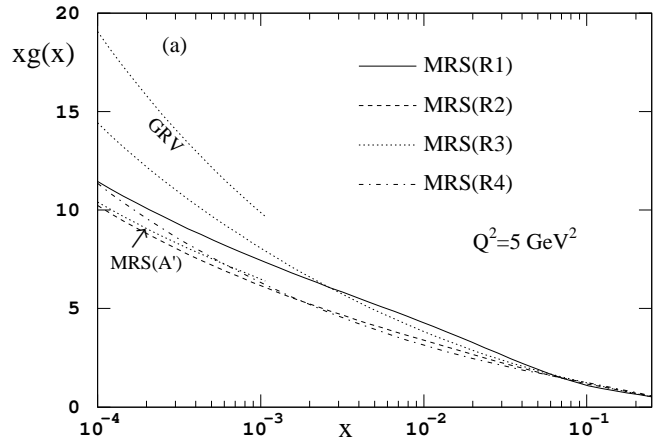,height=6cm}}
\caption{Gluon distributions at $Q^2 = 5\; \GeV^2$
 from  the latest MRS(R) fits \protect\cite{dick}.}
\label{fig:mrsrgluons}
\end{center}
\end{figure}

For $x \lapproxeq 0.01$, the $Q^2$ dependence of $F_2$ is dominated
by the gluon contribution to the right-hand side of the evolution
equation,
\beq
\frac{\partial F_2(x,Q^2)}{\partial \ln Q^2} \simeq
\frac{\as(Q^2)}{\pi} \sum_q e^2_q\;
  \int^1_x \frac{dy}{y} \left( \frac{x}{y}
\right) P_{qg} \left( \frac{x}{y} \right) yg(y,Q^2)\;  .
\label{slope}
\eeq
The uncertainty in the gluon obtained from the HERA data in this way has
steadily decreased in the last few years. As an example,
Fig.~\ref{fig:h1gluon}
shows the gluon obtained from the H1 NLO QCD fit~\cite{h1nlofit}.
The shaded band represents the combined systematic and statistical error
-- approximately $\pm 20\%$ over the range in $x$ ($10^{-4} - 10^{-2}$)
where there is sensitivity. It is interesting to compare the gluon
of Fig.~\ref{fig:h1gluon} with that obtained from a global fit.
Fig.~\ref{fig:mrsrgluons} shows the
gluons at $5\; \GeV^2$ from the latest MRS(R)
series of fits~\cite{dick}, together with several `old' gluons.
At small $x$ ($\sim 10^{-4}$) there is good agreement with the H1 gluon,
and the spread of the new MRS gluons is similar to the uncertainty band
in Fig.~\ref{fig:h1gluon}. However for $x \gapproxeq 10^{-2}$ the
MRS gluon distributions are systematically  smaller than the H1 gluon.
This is due to the extra constraints (principally from momentum
conservation and from prompt photon production) imposed in the global fit.

\begin{figure}[htbp]
\begin{center}
\mbox{\epsfig{figure=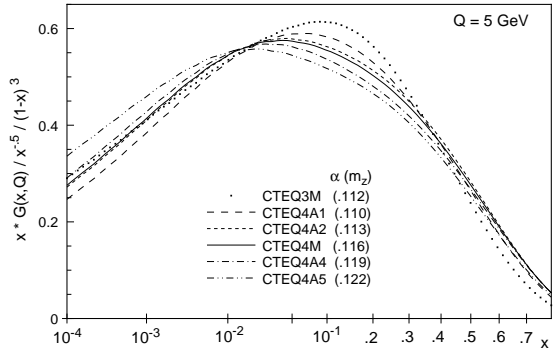,height=5cm}}
\caption{New CTEQ gluon distributions at $5\; \GeV^2$ from fits
incorporating CDF jet data\protect\cite{wuki}.}
\label{fig:wukigluon}
\end{center}
\end{figure}

There has been considerable discussion at this Workshop about the new
large $E_T$ jet data from CDF and D0~\cite{jets}. It is not yet clear
whether the apparent excess of data over theory seen in the CDF data
for $E_T \gapproxeq 200\; \GeV$ is a real effect, and if so whether it
is due to harder parton distributions than previously estimated
\cite{wuki,alan} or to some `new physics' contribution.
What {\it is} clear is that the jet data for $E_T \lapproxeq 200\; \GeV$
have considerable potential for constraining the gluon distribution
($qg \to qg$ is the dominant subprocess) and $\as$. The CTEQ
collaboration have already incorporated the CDF jet data into their
global fits~\cite{wuki}. As an example, Fig.~\ref{fig:wukigluon} shows
a series of CTEQ gluon distributions obtained assuming different
$\as(M_Z^2)$ values in the range $[0.105,0.122]$. The effect of the
jet data is seen in the difference between the new gluons and the
CTEQ(3M) gluon in the $ x \sim 0.1 - 0.3$ range. Notice
also the correlation with the $\as$ value. From such analyses, one may
conclude that the uncertainty in the gluon~\footnote{We refer here to the
uncertainty at the starting scale, i.e. $Q_0^2 \sim 4-5\; \GeV^2$.
Perturbative evolution tends to make the distributions converge
at higher $Q^2$.}
 has now been reduced to
$O(\pm 10\%)$ over a broad range in $x$, rising to
$O(\pm 20\%)$ at small $x \sim 10^{-4}$. This represents considerable
progress!

\begin{figure}[htb]
\begin{center}
\hspace{-3truecm}
\mbox{\epsfig{figure=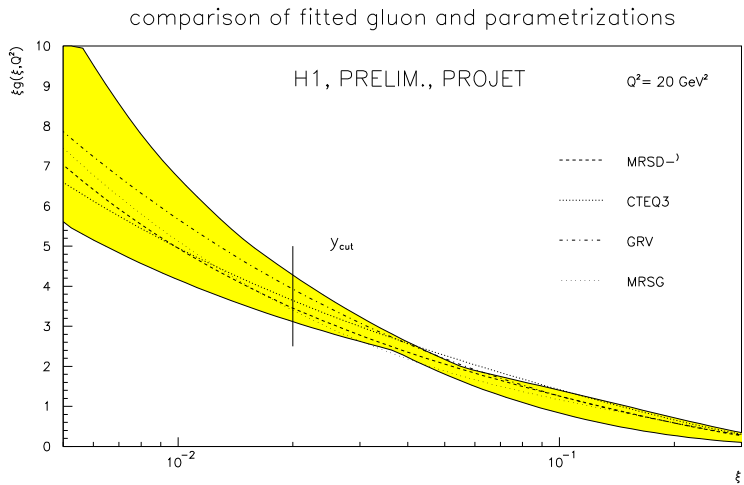,height=6cm,angle=270}}
\vspace{-3.5truecm}
\caption{H1 $2+1$ jet gluon distribution \protect\cite{rosenbauer}.}
\label{fig:h1jet}
\end{center}
\end{figure}

The gluon distribution can also be determined at HERA from the 
`$2+1$' jet cross section, i.e. the cross section for producing
an additional hard hadronic jet in addition to the current and beam
remnant jets. Preliminary results had already  been reported in Paris last year,
but what is new this year is that the analysis can now be done
consistently to next-to-leading order in perturbation theory. The key 
here is the recent availability of  NLO QCD Monte Carlo programs
such as PROJET~\cite{projet} and MEPJET~\cite{mepjet}.
At leading order the cross section is, schematically,
\beq
\sigma(2+1) = \sum_q \left[ q\otimes \hat\sigma(\gamma^* q \to q g) \; + \; 
g \otimes \hat\sigma(\gamma^* g \to q \bar q)\right] \; .
\label{eq:2plus1}
\eeq
Note that the subprocess cross sections are $O(\as)$, and in fact
this allows a reasonably precise determination of the strong coupling,
see Section~\ref{sec:alphas}. The parton distributions
in (\ref{eq:2plus1}) are probed at $ x = x_{\rm Bj}(1+M^2_{jj}/Q^2)$, where
$M^2_{jj}$ is the invariant mass of the final-state
jet pair. In practice, the range covered at HERA is approximately
$0.01 \lapproxeq x \lapproxeq 0.1$. The importance of this is that 
it `fills in' the gap between the small-$x$ scaling violation
and large-$x$ prompt photon methods for determining the gluon.
Both the H1~\cite{rosenbauer}
 and ZEUS~\cite{repond} collaborations have reported
new results on the gluon distribution from $\sigma(2+1)$ 
at this Workshop. Fig.~\ref{fig:h1jet} shows the gluon 
distribution at $20\; \GeV^2$ extracted by H1. The band corresponds 
to the combined systematic and statistical error. Note that there
is good agreement with the standard gluons obtained from global fits.
With a further reduction in the statistical and systematic errors,
this method of determining the gluon will play an important role
in the overall constraint picture.

\begin{figure}[htbp]
\begin{center}
\mbox{\epsfig{figure=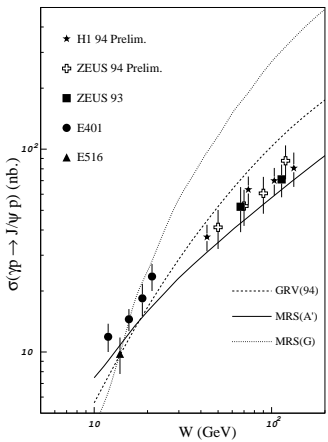,height=6cm}}
\caption{Diffractive $J/\psi$ photoproduction cross section data compared 
with theoretical predictions based on various gluon distributions 
\protect\cite{alanjpsi}.}
\label{fig:jpsigluon}
\end{center}
\end{figure}

Finally, we should mention also a new method of determining the gluon
distribution from diffractive ${\rm J}/\psi$
electro- (and photo-) production: ${\rm e} + {\rm p}
\to {\rm e} + {\rm J}/\psi + {\rm p}$. For sufficiently high $W^2$
the scattering amplitude can be calculated perturbatively~\cite{ryskin}, reducing
(at lowest order) to the scattering of the $c \bar c$ system off the proton
via two gluon colour-singlet exchange. The lowest-order result is
\beq
{d\sigma \over dt}(\gamma^* p \to J/\psi p) = {\Gamma_{ee} M_{\psi}^2 \pi^3
\as(\oQ^2)^2 \over 48\alpha \oQ^6}\; \left[\bar x g(\bar x,\oQ^2)\right]^2
\; \left( 1 + {Q^2\over M_\psi^2} \right) \; ,
\eeq
where
\beq
\oQ = (Q^2 + M_\psi^2)/4\; , \qquad \bar{x} =4 \oQ^2/W^2 \; .
\eeq
At higher orders, and when $\bar{x}$ is small, two gluon exchange
is replaced by a generalized `BFKL' gluon ladder, i.e. the unintegrated
gluon distribution $\cG(x,k_T^2)$ discussed in Section~\ref{subsec:bfkl},
but the basic structure remains the same~\cite{rrml}.
Phenomenological studies based on the above theoretical approach
have recently been performed~\cite{rrml,frankstrik}. The $W^2$ dependence
of the cross section is a direct measure of the shape of the gluon distribution,
and the sensitivity is enhanced by the fact that it enters squared in the
expression for the cross section. Fig.~\ref{fig:jpsigluon}~\cite{alanjpsi} 
compares the  predictions based on various gluon distributions
with recent fixed-target and HERA data. Note that the theoretical
calculations do not yet include the complete $O(\as)$ corrections to the 
cross sections, and cannot therefore distinguish gluons corresponding
to different factorization schemes. In view of the potential of this method
for constraining the gluon at small $x$, the calculation of the NLO
corrections would  be  very worthwhile.

\subsection{High $Q^2$}
\label{subsec:hiq}
\begin{figure}[htb]
\begin{center}
\mbox{\epsfig{figure=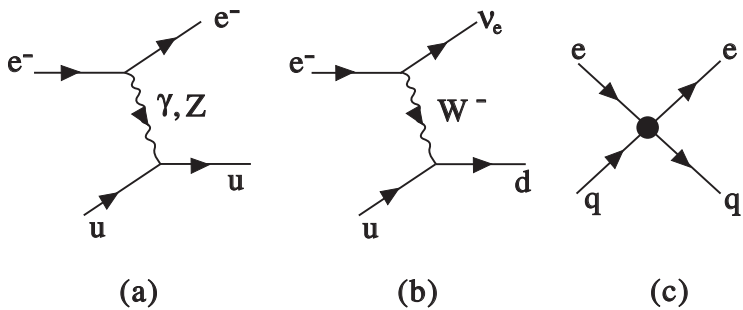,height=4cm}}
\caption{(a) Neutral current, (b) charged current and (c) contact interaction
contributions to electron--quark scattering at high $Q^2$.}
\label{fig:wex}
\end{center}
\end{figure}

\begin{figure}[htb]
\begin{center}
\mbox{\epsfig{figure=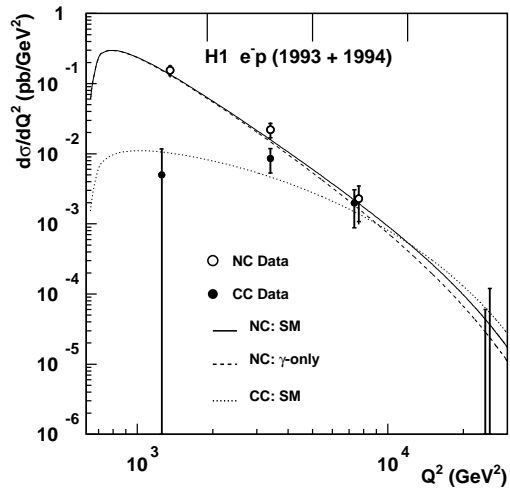,height=7cm}}
\caption{Charged and neutral current deep inelastic cross sections from H1 at
HERA \protect\cite{h1nccc}.}
\label{fig:ccnc}
\end{center}
\end{figure}

\begin{figure}[htb]
\begin{center}
\mbox{\epsfig{figure=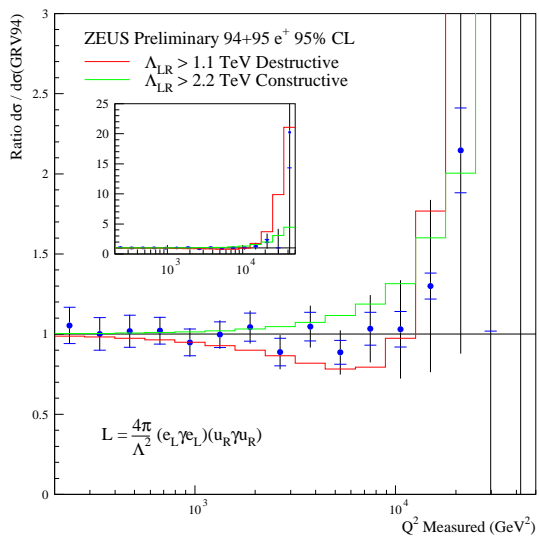,height=8cm}}
\caption{Limits on lepton--quark contact interactions from high-$Q^2$
neutral current DIS data, from ZEUS \protect\cite{zeussum}.}
\label{contact}
\end{center}
\end{figure}

Over most of the measured $Q^2$ range in Fig.~\ref{phase}, the deep inelastic
cross section is dominated by the neutral current contribution,
Fig.~\ref{fig:wex}(a).
With increased statistics at high $Q^2$ ($\gapproxeq 10^4\; \GeV^2$), 
HERA can also measure the  structure of the proton as seen by the
{\it charged} weak current, Fig.~\ref{fig:wex}(b). The corresponding 
 deep inelastic cross section  is
\beq
{d^2\sigma^{CC} \over d x d Q^2} =   {(1-\lambda_e ) \pi\alpha^2 \over
8 \sin^4\theta_W  ( Q^2 + M_W^2)^2 }
 \sum_{i,j} 
\left[  \vert V_{u_id_j}\vert^2  u_i(x)  
  + (1-y)^2  \vert V_{u_j d_i}\vert^2  d_i(x)   \right]  ,
\eeq
where $\lambda_e$ is the helicity of the electron, $u_i$ and $d_i$ refer
to up- and down-type quarks respectively,  and the $V_{u_id_j}$
are the elements of the CKM matrix. In principle this allows
a different combination of parton distributions to be measured.
Some preliminary results have been reported at this Workshop~\cite{h1sum}.
The data clearly show the convergence
of the charged and neutral current cross sections at very high $Q^2$,
$d\sigma^{CC} / d x d Q^2 \sim d\sigma^{NC} / d x d Q^2 \sim Q^{-4}$,
see Fig.~\ref{fig:ccnc}. The fact that the high-$Q^2$ cross sections agree
with the DGLAP--based extrapolations enables limits to be placed on possible
contact interactions, as expected on general grounds in composite models.
The convention (see for example~\cite{ehlq})
is to parametrize such interactions using the Lagrangian
\beq
\Delta {\cal L} = {4\pi \eta \over \Lambda^2}\; \sum_{q,l}
({\bar q}\gamma^{\mu} q)_{L,R}\; ({\bar l}\gamma_{\mu} l)_{L,R}\; ,
\eeq
which leads to additional contributions to the cross section
 of order $Q^{-2}\Lambda^{-2}$ and $\Lambda^{-4}$. The parameter $\Lambda$
is a measure of the compositeness scale, and $\eta$ determines whether
the interference with the Standard Model cross section is constructive or
destructive. An example of how the data can be used to constrain $\Lambda$
is shown in Fig.~\ref{contact} from ZEUS~\cite{zeussum}. 
Here a particular (left--right)
 helicity structure has been assumed for the lepton--quark contact interaction.
The actual lower limits on $\Lambda$ depend slightly on this structure,
and are typically $O(1\; \TeV)$ and $O(2\; \TeV)$ for destructive and
constructive interference respectively. Note that somewhat higher
 limits ($3-4\; \TeV$) are obtained
from  Drell-Yan ($q \bar q \to l^+l^-$) lepton-pair production at the Fermilab 
Tevatron collider~\cite{bodek}.

\section{$\as$ measurements}
\label{sec:alphas}
Deep inelastic scattering has traditionally provided some of 
the most precise measurements
of the strong coupling $\as$. There are three basic methods:
\begin{itemize}
\item scaling violations of structure functions,
\item structure function sum rules,
\item jet cross sections.
\end{itemize}
Some of the high-precision values~\footnote{Throughout this section, $\as$ refers
to $\as(M_Z^2)$ in the $\overline{\mbox{MS}}$ renormalization
scheme with $n_f = 5$.}
 obtained using these methods are listed in Table~\ref{disalphas}.
\begin{table}[htb]
\label{disalphas}
\begin{center}
\begin{tabular}{|lllll|}  \hline
\rule[-1.2ex]{0mm}{4ex}
measurement & experiment & $\langle Q\rangle$ & pQCD &
$\as^{\msb}(M_Z^2)$ \\ \hline
\rule[-1.2ex]{0mm}{4ex} $\partial F_2^{lN} /\partial\ln Q^2$
& SLAC/BCDMS~\cite{ALAIN} & $7.1$ & NLO &  $0.113 \pm 0.005$  \\
\rule[-1.2ex]{0mm}{4ex}  $\partial F_2^{\nu N} /\partial\ln Q^2$
& CCFR~\cite{ccfrsv}   & $5.0$ & NLO &  $0.111 \pm 0.006$  \\
\rule[-1.2ex]{0mm}{4ex} $\int_0^1 dx (F_3^{\nu N} + F_3^{\bar\nu N})$
& CCFR~\cite{ccfrgls}   & $1.73$ &  NNLO & $0.108^{+0.006}_{-0.009}$  \\
\rule[-1.2ex]{0mm}{4ex} $\int_0^1 dx (g_1^{p} - g_1^{n})$
& SLAC/EMC/SMC~\cite{elka}   & $1.58$ &  NNLO & $0.122^{+0.005}_{-0.009}$  \\
\rule[-1.2ex]{0mm}{4ex} $\sigma^{\gamma^* p}(2+1\; \mbox{jet})$
& H1, ZEUS~\cite{oldjet}   & $22.1$ &  NLO & $0.118\pm 0.009$  \\
\hline
\end{tabular}
\caption{Some deep inelastic scattering $\as$ measurements.}
\end{center}
\end{table}
The errors are between 5\% and 10\%, with the highest accuracy
coming from the scale variation of the structure functions.
More importantly, all the results agree within errors. These `deep inelastic'
$\as$ measurements also have a significant weight in the overall
world average value, which is currently~\cite{book}
\beq
\as(M_Z^2) = 0.116 \pm 0.005 \; .
\eeq

\begin{figure}[htbp]
\begin{center}
\mbox{\epsfig{figure=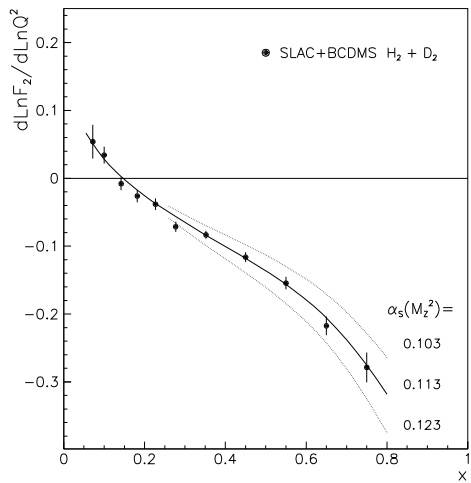,height=7cm}}
\caption{Logarithmic $Q^2$ derivative of the $F_2$ structure function
measured off hydrogen and deuterium targets compared to NLO 
predictions, from Ref.~\protect\cite{ALAIN}.}
\label{fig:mimile}
\end{center}
\end{figure}

The value of $\as$ obtained from scaling violations has been discussed
widely at this Workshop, especially in connection with new information
coming from the HERA $F_2$ data. It is important to recall that the 
`gold-plated' values in Table~\ref{disalphas} come from the high-precision
fixed-target data. In particular, in the analysis of Ref.~\cite{ALAIN},
which includes lower energy SLAC data and a
phenomenological contribution from higher-twists, the resolving power
on $\as$ comes from data with $0.3 \lapproxeq x \lapproxeq 0.5$ 
(see Fig.~\ref{fig:mimile}) and $10 \lapproxeq Q^2 \lapproxeq 100\; \GeV^2$.
In this kinematic region, higher-order perturbative and higher-twist
corrections should be very small. 

In the last year it has become clear that the HERA structure function
data prefer a slightly larger $\as$ value than the 
fixed-target data at higher $x$, a result first pointed
out in Ref.~\cite{mrsg}.  For example, in the NLO analysis
of the 1993 data reported in Ref.~\cite{bfas}, the value
$\as(M_Z^2)  = 0.120 \pm 0.005({\rm exp.}) \pm 0.009({\rm th.})$ 
was obtained (for an update
see Ref.~\cite{ball}),
with the theory error dominated by the variation with respect to the
renormalization and factorization scale. In the new MRS analysis~\cite{dick},
which includes the new 1994 HERA data, fits are performed with two $\as$
values, $0.113$ and $0.120$. The latter value is motivated by (a) the
preference of the LEP/SLC three-jet and total $Z$ hadronic width data for
a larger $\as$ value (see for example Ref.~\cite{book}), and (b)
the fact that the Fermilab inclusive medium--$E_T$ jet data is slightly 
steeper than the theoretical prediction using partons with $\as \simeq
0.113$~\cite{dick,wuki}. 
Table~\ref{herachisq} shows the $\chi^2$ values for the MRS 
fits~\cite{dick} to the 
H1 and ZEUS data with the two different $\as$ values. Both experiments
show a clear preference for the larger value.
\begin{table}[htb]
\label{herachisq}
\begin{center}
\begin{tabular}{|l|c|c|}  \hline
\rule[-1.2ex]{0mm}{4ex}
expt.  & $\chi^2({\rm R}_1)$  & $\chi^2({\rm R}_2)$  \\ 
\rule[-1.2ex]{0mm}{4ex}
  & $\as = 0.113$  & $\as = 0.120$  \\ 
\hline
H1 (193 pts)  & 182  & 168 \\
ZEUS (210 pts)  & 391  & 362 \\
\hline
\end{tabular}
\caption{$\chi^2$ values for the new MRS(R$_1$,R$_2$) fits to the HERA $F_2$
data \protect\cite{dick}.}
\end{center}
\end{table}

What could be responsible for this apparent difference in the 
small- and large-$x$ $\as$ values (if one ignores the fact that 
within the overall errors there is no significant disagreement!)?
So far there has been no proper study of the dependence of the 
$\as$ value on the way the charm contribution to $F_2$ is treated.
This is almost certainly insignificant at large $x$, but could
make a slight difference at small $x$ where the charm contribution
is expected to be a more significant fraction of the total structure
function, and where the threshold $Q^2$ dependence of $F_2^c$
could perhaps `fake' a stronger DGLAP $Q^2$ variation.
Second, we have already seen that higher-order contributions
to DGLAP evolution become more and more important as $x \to 0$.
It could well be that the larger $\as$ value at small $x$ is an 
indication that the (as yet unknown) NNLO corrections are not
negligible.  A complete calculation of the third-order splitting functions
 would almost certainly settle this issue.
 
 New analyses of $\as$ from the deep inelastic $2+1$ jet cross section
(Eq.~(\ref{eq:2plus1})) by H1 and ZEUS have also been reported at this 
Workshop~\cite{rosenbauer,h1as}.
It appears that there may have been a slight problem with the
previous analyses~\cite{oldjet} arising from  a mismatch of the 
jet algorithm
used in the NLO theoretical calculation  and that employed in the 
experimental analyses. In general, jets are defined using
a cluster algorithm which is applied to hadrons and partons
in the experimental and theoretical analyses respectively.
One defines a dimensionless `metric' between two hadrons (say),
$y_{ij} = s_{ij} / M^2$,
and merges the hadrons into a single cluster if
 $y_{ij} < \ycut$, where $\ycut$ is a fixed (small) number.
 Various choices for $s_{ij}$ ($ = 2E_i E_j (1-\cos\theta_{ij})$
 [JADE], $ = {\rm min}\,k_{Tij}^2$ [Durham/$k_T$], \ldots)
 and $M^2$ ($ = Q^2, W^2,\ldots$) are possible, but in each case
 {\it the theoretical calculation must be matched to the experimental
 definition} -- the NLO corrections are in general different 
 for the different jet definitions.
The new MEPJET theory Monte Carlo~\cite{mepjet} allows an arbitrary
algorithm to be used, and can therefore be  tuned to the experimental
analysis. Although there are still some questions to be answered
(the effect of hadronization, the correlation of $\as$ with the
gluon distribution, the sensitivity to the jet cuts, etc.), it is already
clear that this is a potentially powerful method for measuring
the strong coupling at HERA in the future. It will
 be very interesting  to see if the value thus obtained agrees with the
 scaling violation value.~\footnote{There is an analogy here
 with the two methods (total hadronic width and jet rates) for measuring
 $\as$ at LEP/SLC.}

\section{Spin physics}
\label{sec:spin}
Polarized deep inelastic scattering experiments provide information
about how the spin of the hadron is shared among its parton constituents.
As for unpolarized scattering,  the cross section can be related
to two structure functions, $g_1$ and $g_2$. The former is analogous
to $F_1$, and in the parton model is given by a sum over polarized
quark distribution functions:
\beq
g_1(x) = \half \sum_q e_q^2 [\Delta q(x) + \Delta \bar{q}(x)]\; .
\eeq
with $\Delta q = q^\uparrow - q^\downarrow$.
In perturbative QCD, the distributions $\Delta q$ and $\Delta g$ acquire
a scale dependence given by the polarized versions of the DGLAP equations
($\Delta$DGLAP).
The first moments, $\int_0^1dx \Delta q_i$ and $\int_0^1 dx \Delta g$,
represent the net spin carried by the various partons. The former
can be related to axial-vector couplings measured in hyperon decay.
In addition, the  Bjorken sum rule ($\int_0^1 dx (g_1^p - g_1^n) = 
{1\over 6} \vert {g_A/g_V}\vert [1+O(\as)]$) provides a fundamental
test of the theory.

\begin{figure}[htb]
\begin{center}
\mbox{\epsfig{figure=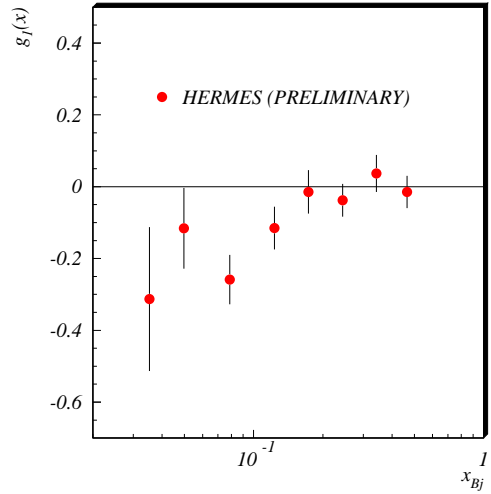,height=9cm}}
\caption{The spin structure function $g_1^n(x)$ from HERMES 
\protect\cite{hermes}.}
\label{fig:hermes}
\end{center}
\end{figure}

There are currently several important theoretical issues:

\begin{itemize}
\item How much of the spin is carried by each type of parton; in particular,
how much is carried by strange quarks and gluons?
\item Is the observed $Q^2$ dependence of the structure functions consistent
with $\Delta$DGLAP evolution?
\item How large is $g_2$ and is it consistent with model calculations?
\item Can the theory predict the $x\to 0$ behaviour of the polarized
distributions, as for unpolarized distributions
 (see Section~\ref{subsec:bfkl})?
\end{itemize}
Answers to all of these questions require precision input from experiments.
The last few years have seen a rapid growth in the amount of data available.
At this Workshop, the SMC collaboration~\cite{smc} have reported
new measurements of $g_1^d$, $g_2$ and the semi-inclusive 
process $\mu N \to N h X$. The SLAC E-143 collaboration~\cite{e143}
have reported new measurements of $g_2$ and the $Q^2$ dependence of
 $g_1^{p,d}$. But perhaps the most significant development has been
 the first results from the HERMES experiment at DESY~\cite{hermes}.
Fig.~\ref{fig:hermes} shows preliminary results on $g_1^n$.
The results are in agreement 
with published SLAC-E142 measurements~\cite{e142} which cover approximately
the same kinematic range. One of the advantages of the HERMES experiment
is the ability to measure semi-inclusive processes, i.e. 
$e^{\updn} p^{\updn}   \to  e + h + X$. The cross section for this
 is, schematically,
\beq
\Delta \sigma^h(x,z) \sim \sum_q \Delta q(x) D_{\Delta q \to h}(z)\; .
\eeq
The identification of  the charge and flavour of hadrons in the final state
therefore gives a more powerful handle on the various $\Delta q$, in particular
the separation between valence and sea quarks. Quantitative results
from HERMES are expected soon.

\begin{figure}[htbp]
\begin{center}
\mbox{\epsfig{figure=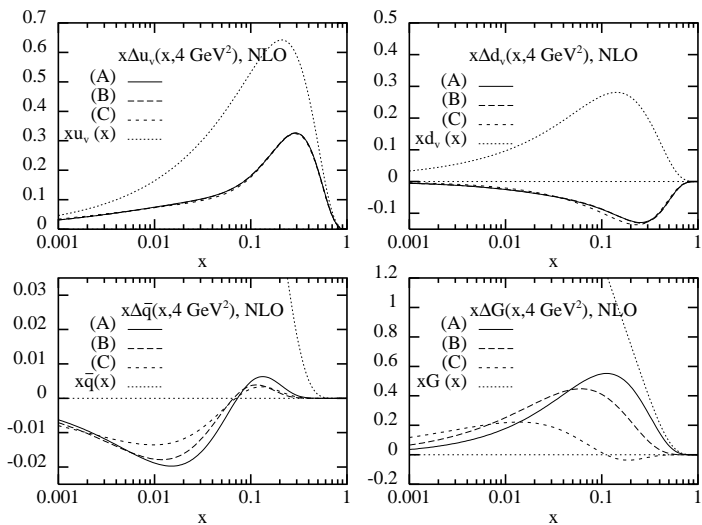,height=7cm}}
\caption{Polarized quark and gluon distributions obtained
from a  NLO QCD fit \protect\cite{gsnlo}.}
\label{fig:gsnlo}
\end{center}
\end{figure}

New theoretical developments have also been reported at this Workshop.
The major advance in the last year has been the calculation
of the two-loop ($O(\as^2)$) polarized splitting functions~\cite{pol_nlo},
\beq
\Delta P_{ab}(x,\as) = \as \Delta P_{ab}^{(0)}(x)
+ \as^2 \Delta P_{ab}^{(1)}(x) + \ldots\; .
\eeq
This allows a consistent NLO $\Delta$DGLAP analysis of the polarized
structure function data.
 Already several groups have performed global fits
to determine the parton distributions $\Delta q_i$ and $\Delta g$ 
\cite{grsv,bfr,gsnlo}. 
While the data can indeed be described by  a simple set of starting 
distributions, there is not yet sufficient experimental
information to determine the full quark flavour structure. With
data on $g_1^p$ and $g_1^n$, only the valence $\Delta u$ and $\Delta d$ quark
distributions are well constrained. Further assumptions have 
to be made to extract the ($\Delta q_S$) sea quark and gluon distributions
(for example, SU(3) flavour symmetry for the sea, and Regge behaviour
for the $x\to 0$ behaviour of  $\Delta q_S$ and $\Delta g$).
As for the unpolarized distributions, weak information on the
gluon distribution is obtained from the $Q^2$ dependence of the structure
functions.  Fig.~\ref{fig:gsnlo} shows an example of polarized distributions
at $Q_0^2 = 4\; \GeV^2$ obtained from a global fit~\cite{gsnlo}.
The three sets of lines labelled (A), (B), and (C) 
correspond to fits with different assumptions about the form of $\Delta g$
at large $x$.  

For the future, more data are urgently  needed to further constrain
the parton distributions. Improved precision on the $Q^2$ dependence
of $g_1$  at small $x$ will help to  determine $\Delta g$ and the semi-inclusive
measurements of $\pi^\pm$ and $K^-$ will constrain $\Delta q_V$ and $\Delta s$
respectively. The measurement of $\Delta g$ at medium and large $x$ presents a real
challenge. As for the unpolarized distribution, what is needed is a process
in which the gluon enters at lowest order. In this context, processes such
as
\beqn
{\rm e}^{\updn} {\rm p}^{\updn} &  \to & e \; +\;  \mbox{jets}\; + \; X \; ,
\nonumber \\
\gamma^{\updn}{\rm  p}^{\updn} &  \to & {\rm J}/\psi \;(\mbox{or}\; c \bar c)
\; + \; X\; , \nonumber \\
{\rm p}^{\updn} {\rm  p}^{\updn} &  \to & (\gamma, \; \mbox{jets}, \; \dots )
\; + \; X  \nonumber
\eeqn
have been discussed at the Workshop, 
see for example Ref.~\cite{polmirkes}. An important point to
remember is that for any such process, one must be sure that the
collision energies, luminosities, cuts etc. are high enough that the 
{\it unpolarized} gluon distribution can be reliably extracted from
the corresponding unpolarized scattering process.

Finally, the behaviour of $g_1$ as $x\to 0$ has been discussed, in the context
of a `BFKL-type' resummation of large $\ln(1/x)$ 
logarithms~\cite{ermolaev_ns,blumlein_ns,ermolaev}. As for the unpolarized
(singlet) structure function, the behaviour of (singlet) $g_1^S$ at small $x$
is controlled by the $t$-channel exchange of a generalized gluon ladder.
However there are important qualitative differences in the resulting
behaviour. In the unpolarized case, both $t$-channel gluons are longitudinally
polarized and the resulting behaviour is $F_2^S(x,Q^2) \sim x g \sim (1/x)^{O(\as)}
\sim (1/x)^{0.5}$, as shown in Section~\ref{subsec:bfkl}. In the polarized case,
this leading polarization configuration cancels, and the leading behaviour
has one gluon transversely polarized. The predicted small-$x$ behaviour,
obtained from resumming the terms proportional to $[\as \ln^2(1/x)]^n$
to all orders, is now the less singular  $g_1^S(x,Q^2) \sim (1/x)^{O(\sqrt{\as})}$.
A detailed calculation~\cite{ermolaev} gives a value $\lambda  = 3.45
\sqrt{N\as/(2\pi)} \approx 1.0$ for the power of $1/x$. It will be interesting to 
see whether such behaviour can  be measured experimentally. 
The same calculation gives the leading $\ln(1/x)$ contributions to the 
polarized splitting functions at all orders in perturbation theory, cf.
Section~\ref{subsec:beyondnlo}. The impact of these on the standard
$\Delta$DGLAP evolution at small $x$ has been studied in Ref.~\cite{blumlein}.
As for the unpolarized case, it is found that, in practice, 
subleading logarithms are likely to play an important role.

\section{Photoproduction}
\label{sec:photo}

The high collision energies available at HERA
allow photoproduction processes to be studied at very short 
distances. Hard scattering processes such as jet and charm production
yield tests of perturbative QCD at next-to-leading order and provide
information about the quark and gluon structure of the photon.
Since the very first measurements of photoproduction cross sections
at HERA in 1993, progress in this area has been steady. The new results
of the past year are summarized in the report of the Photoproduction
Working Group in these Proceedings. Here I will briefly mention
two of these results which are particularly interesting.

Large $E_T$ dijet photoproduction proceeds at leading order
via  the `direct' scattering processes 
$\hat\sigma(\gamma q \to q g)\otimes q^p$ and
$\hat\sigma(\gamma g \to q \bar q)\otimes g^p$, and at higher orders through
the `resolved' processes
 $\hat\sigma(q q \to q q)\otimes q^\gamma\otimes q^p$,
 $\hat\sigma(g q \to g q)\otimes g^\gamma\otimes q^p$, etc.
The latter are particularly important at small $E_T$.
At leading order, 
the momentum fraction of the photon constituent can be reconstructed
from the transverse energies and rapidities of the dijets, 
\beq
x_\gamma  = {E_{T1} \exp(-\eta_1) + E_{T2} \exp(-\eta_2)
  \over 2 E_\gamma }\; .
\eeq
Thus direct scattering events have $x_\gamma \approx 1$ and resolved scattering
events have $x_\gamma < 1$. 

\begin{figure}[htbp]
\begin{center}
\mbox{\epsfig{figure=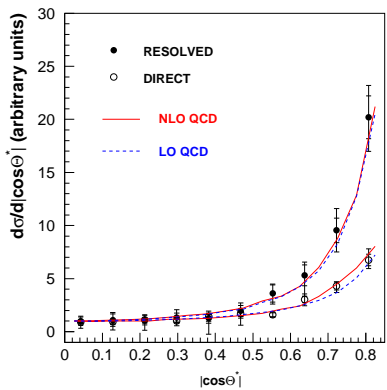,height=6cm}}
\caption{Dijet angular distribution, from Ref.~\protect\cite{feld}.}
\label{fig:zeusdijet}
\end{center}
\end{figure}

The angular distribution of the dijets probes the structure of
the hard scattering process. In particular, small angle scattering
is sensitive to the spin of the $t$-channel exchange particle.
In this respect there is an important difference between the direct
and resolved scattering cross sections, which  are mediated by spin$-\half$
(quark) 
and spin$-1$ (gluon) exchange respectively. Thus 
\beqn
{d \sigma^{\rm dir} \over d \cos\theta^*} &\sim &
g^p \otimes \hat\sigma(\gamma g \to q \bar q) + \ldots\ \longrightarrow\
{1 \over 1 - \vert \cos\theta^* \vert }      \nonumber \\
{d \sigma^{\rm res} \over d \cos\theta^*} &\sim &
q^\gamma \otimes q^p \otimes \hat\sigma(q q  \to q q) + \ldots
\ \longrightarrow\
{1 \over\left( 1 - \vert \cos\theta^* \vert\right)^2 }\; ,
\eeqn
where $\theta^*$ is the scattering angle in the dijet centre-of-mass frame.
In Fig.~\ref{fig:zeusdijet}, from the ZEUS collaboration~\cite{feld}, 
a cut in $x_\gamma$ $(0.75)$ 
is used to separate the events into the two `direct' and `resolved' classes.
The expected difference in the distribution as $\vert\cos\theta^*\vert
\to 1$ between the  `direct' and `resolved' samples is clearly seen,
and in both cases there is good agreement with NLO QCD predictions.

\begin{figure}[htbp]
\begin{center}
\mbox{\epsfig{figure=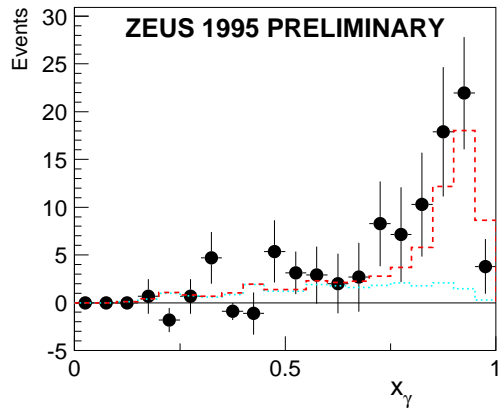,height=6cm}}
\caption{Distribution in $x_\gamma$ of prompt photon events at HERA,
 from Ref.~\protect\cite{bussey}.}
\label{fig:bussey}
\end{center}
\end{figure}

Another interesting result concerns the first quantitative measurements at HERA
of large $p_T$ photon photoproduction~\cite{bussey}. At leading order,
this proceeds via the Compton scattering of a photon off a quark:
\beq
d \sigma\;  \sim\;  \sum_q e_q^4     \; q^p \otimes
 \hat\sigma(\gamma q \to \gamma  q) + \ldots \; .
\eeq
At higher orders there are also contributions from resolved processes
such as $qg \to \gamma q$. Fig.~\ref{fig:bussey} shows the 
distribution in $x_\gamma$ for direct photon events from the 
ZEUS collaboration. A clear excess at large $x_\gamma$
is seen, signalling the direct Compton scattering process. 
The interesting feature of the Compton process
is that it is proportional to the {\it fourth} power of 
the quark charge, and therefore probes a different linear combination
of quark distributions in the proton than the deep inelastic structure 
function~\cite{bawa}.
The photon angular distribution of the Compton process should again
exhibit a characteristic $(1 - \vert \cos\theta^* \vert)^{-1} $ form
at small angles.

\section{Diffractive deep inelastic scattering}
\label{sec:harddiff}

The observation of `diffractive' 
deep inelastic events with a large rapidity gap~\cite{gaporig}
 has spawned a new field
of interest at HERA in the last few years. At this Workshop, as last year,
there was much lively discussion about the measurement and interpretation
of these events. 

The focus of attention so far is the diffractive structure function
$F_2^D$. This is obtained from a subclass of DIS events in which
there is a large rapidity gap (H1, ZEUS) or from the observed excess
of events with small $M_X$ (the invariant mass of hadrons in the detector)
 over the expected non-exponentially-suppressed 
non-diffractive contribution (ZEUS). The ZEUS collaboration have also identified
energetic, small-angle protons in  the Leading Proton Spectrometer (LPS),
 consistent
with the interpretation that  $e p \to e p X$ is the underlying process.

\begin{figure}[htbp]
\begin{center}
\mbox{\epsfig{figure=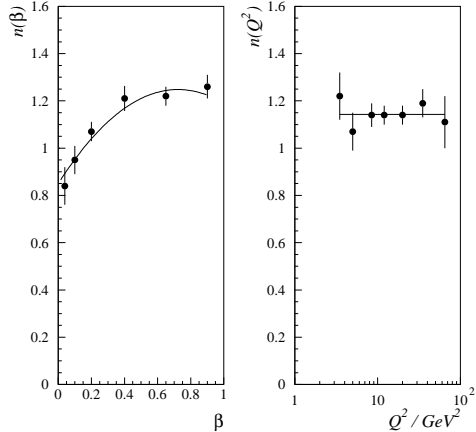,height=6cm}}
\caption{Measurement  of the parameter $n$, where $F_2^{D(3)}
\sim x_p^{-n}$, from H1~\protect\cite{H1thisyear}.}
\label{fig:h1beta}
\end{center}
\end{figure}

The forward proton momentum can be parametrized in terms
of the (small) momentum transfer $t$ and the longitudinal 
momentum fraction $1-x_p \approx 1$. One can then  define a diffractive
structure function $F_2^{D(4)}(\xbj,Q^2; x_p,t)$ or equivalently
$F_2^{D(4)}(\beta,Q^2; x_p,t)$ where $\beta = \xbj/x_p 
\simeq Q^2/(Q^2 + M_X^2)$. If $t$  is unmeasured and therefore integrated
over, one can define a corresponding $F_2^{D(3)}(\beta,Q^2; x_p)$ function.
The crucial experimental observations are that
$F_2^{D(4)}$ is a {\it leading twist} structure function, indicating
deep inelastic scattering off point-like objects,  and that 
there is an approximate factorization property:
\beq
F_2^{D(3)} \sim x_p^{-n} {\cal F}(\beta,Q^2)\; .
\eeq
At the Paris Workshop last year, the values $n = 1.19 \pm 0.06({\rm stat.})
\pm 0.07 ({\rm sys.})$ (H1~\cite{H1lastyear})
and $n = 1.30 \pm 0.08({\rm stat.})
{+0.08 \atop -0.14} ({\rm sys.})$ (ZEUS~\cite{ZEUSlastyear}) were reported.
This year, with new data, H1 have measured $n$ as a function 
of $\beta$ and $Q^2$, see Fig.~\ref{fig:h1beta}.
A clear dependence on $\beta$ (but not on $Q^2$) is observed.
ZEUS have also reported new values: $n = 1.46 \pm 0.10$ from 
$F_2^{D(2)}$ defined  by the $M_X$ distribution~\cite{ZEUSthisyearMX},
 and $n = 1.28 \pm 0.17$
from the LPS data~\cite{ZEUSthisyearLPS}. Note that these are
 not in contradiction,
since the $\langle\beta\rangle$ averages are different for the two samples.
However, as last year, they do appear to be slightly larger than the H1 values.

The most attractive theoretical interpretation of these diffractive
DIS events is that they correspond to the structure function of the 
pomeron ($\cP$), rather than the proton (for a review see Ref.~\cite{PVL}).
In its simplest version, this model would predict
\beqn 
 F_2^{D(4)}(\beta,Q^2; x_p,t)& =& f_{\cP}(x_p,t)\;F_2^{\cP}(\beta,Q^2)\; ,
 \nonumber \\
 \quad f_{\cP}(x_p,t) = F_{\cP}(t) x_p^{2\alpha_{\cP}(t)-1}\; ,&&
 F_2^{\cP}(\beta,Q^2) = \beta \sum_q e_q^2 q_{\cP}(\beta,Q^2)\; ,
\label{reggefact}
\eeqn
i.e. a factorized structure function with $n \approx  2\alpha_{\cP}(0)-1
\approx 1.16$. This model is based on the notion of `parton constituents
in the pomeron' first proposed by Ingelman and Schlein~\cite{IS} and supported
by data from UA8. In such a model, a modest amount of factorization breaking
could be accommodated by  invoking a sum over Regge trajectories, each
with a different intercept and structure function:
\beq
 F_2^{D(4)}(\beta,Q^2; x_p,t) = \sum_{\cR}\;
 F_{\cR}(t) x_p^{2\alpha_{\cR}(t)-1}\;F_2^{\cR}(\beta,Q^2)\; ,
\label{eq:reggeons}
\eeq
which would yield an effective $n$ which depends on $\beta$
but is approximately independent of $Q^2$.

\begin{figure}[htbp]
\begin{center}
\mbox{\epsfig{figure=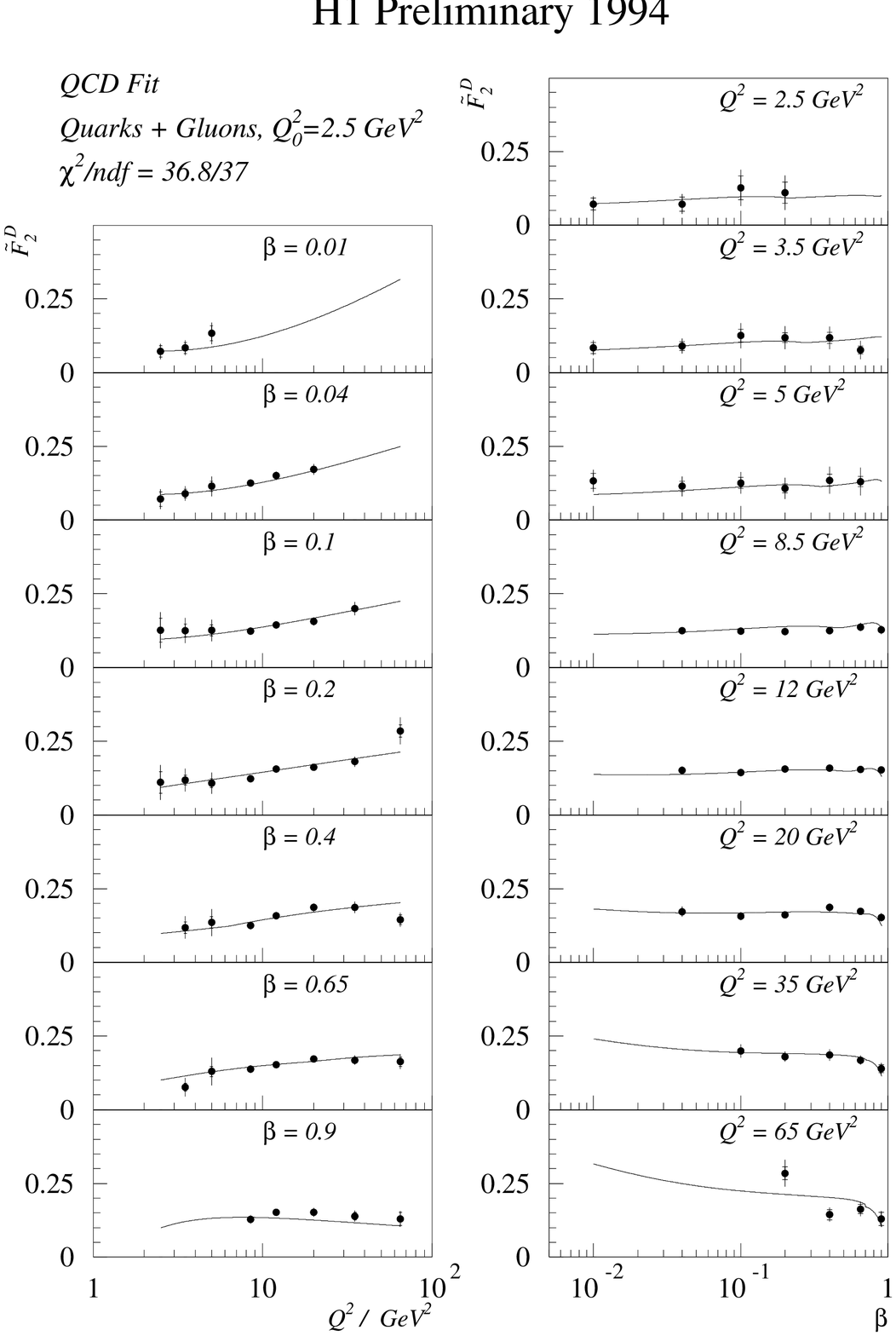,height=8cm}}
\caption{The diffractive structure function  $F_2^{(D)}$ 
measured by H1~\protect\cite{H1thisyear}, together with the result
of a NLO QCD fit.}
\label{fig:h1nlofit}
\end{center}
\end{figure}

\begin{figure}[htbp]
\begin{center}
\mbox{\epsfig{figure=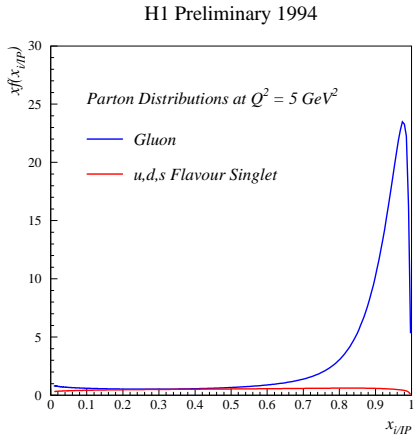,height=6cm}}
\caption{Parton distributions obtained from the fit to the data
shown in Fig.~\protect\ref{fig:h1nlofit}~\protect\cite{H1thisyear}.}
\label{fig:h1partons}
\end{center}
\end{figure}

The above approach has been put on a firmer theoretical footing
in Ref.~\cite{soper}, where the general concept of `diffractive
parton distributions' is introduced. In particular it is shown 
that an operator product definition exists, and that the diffractive
distributions should satisfy DGLAP evolution equations.
There has been a number of analyses based on DGLAP fits to 
diffractive structure function data in the last  year. At 
this Workshop the H1 collaboration has presented a new measurement of
$F_2^{(D)}$ (i.e. integrated over both $x_\cP$ and $t$) and a new NLO QCD fit,
see Fig.~\ref{fig:h1nlofit}.
The resulting parton distributions at $Q^2 = 5\; \GeV^2$
are shown in Fig.~\ref{fig:h1partons}.
Evidently the `pomeron' (more precisely, the aggregate of the colour-singlet
exchanges in Eq.~(\ref{eq:reggeons})) is a predominantly {\it gluonic}
object. Although the structure function only directly measures the 
quark distributions, a very hard gluon distribution is needed to 
give an approximate scaling (in $Q^2$) behaviour at large $\beta$,
see Fig.~\ref{fig:h1nlofit}.

From a purely phenomenological point of view, the above model
has many attractive features. It enables predictions to be made, for
example for the charm content of the diffractive structure
 function~\cite{gspom} and the rate of jet production in diffractive
 DIS events~\cite{H1diffjets}. With a further assumption
 of {\it universality} of diffractive parton distributions, which 
 lacks any theoretical proof at present however, one can make
 quantitative predictions for diffractive hard scattering processes
 in photon-hadron and hadron-hadron collisions~\cite{zeusphoto,pomks}.

Finally, we should mention a different approach to diffractive
deep inelastic scattering based on perturbative QCD. In the 
microscopic colour dipole model~\cite{nikzak}, a generalized `BFKL'
gluon ladder interacts with the $q \bar q$ Fock state
of the virtual photon. In this model many features of
 $F_2^{D}$ can be predicted from QCD perturbation theory.
The most striking features are the breaking of the Regge factorization
of Eq.~(\ref{reggefact}), and the qualitative differences between the 
longitudinal and transverse structure functions~\cite{nikzak}:
\beq
F_T^D \sim g^2\left(x_p,{m_q^2\over 1-\beta}\right)\; ,\quad
F_L^D \sim {1\over Q^2} g^2\left(x_p,{Q^2 \over 4}\right)\; .
\eeq
Note that there is no $x_p - (\beta,Q^2)$ factorization, and also
no DGLAP evolution as $\beta \to 1$ at high $Q^2$, where $F_T^D$
dominates. The $x_p$ dependence  is determined  by the $x\to 0$
behaviour of the gluon distribution, and is generally steeper than the 
behaviour predicted by the simple (soft)  pomeron model. 
For heavy quarks, a perturbatively hard scale is set by $m_q^2$,
and infra-red safe predictions can be made from first principles
~\cite{nikzak,alanjpsi}.
Like the soft pomeron model, the pQCD 
model is so far in good agreement with the data.
Once again the conclusion is obvious: more data, not only on 
$F_2^{D(4)}(\beta,Q^2; x_p,t)$ but also on $F_L^D$, charm and jet
production, are urgently needed to discriminate between these different
approaches.

\section{Conclusions}
\label{sec:conc}
At this Workshop we have seen significant progress in almost all 
aspects of deep inelastic scattering physics. Advances in experimental
measurements are matched by advances in theoretical calculations,
and our understanding of the short-distance structure of hadrons steadily
improves as a result. However, each new Workshop inevitably produces
new questions to be answered. Some of these have been discussed in this summary:
for example, why does NLO DGLAP work so well even at small $x$ and $Q^2$ values,
can we push the uncertainty in the gluon distribution below
the $10\%$ level, is there a systematic difference between the
deep inelastic  $\as$ values extracted from high and low $x$, how can
the polarized gluon distribution be directly measured, and can the 
deep inelastic diffractive data really be understood in a simple 
`pomeron parton model' or will the explicit factorization-breaking
properties of the pQCD model be revealed?
 We look forward to some answers next year!

\section*{Acknowledgments}
The success of DIS96 has been  due in no small measure to the excellent
organization provided by Giulio D'Agostini and his team.
I am very grateful to them for all their help and support before,
during and after the meeting. Thanks also to all those colleagues
who so efficiently provided  results and figures,
 and  to  Keith Ellis,
Thomas Gehrmann, Alan Martin, Dick Roberts
and Bryan Webber for enjoyable collaborations, some of the fruits of which
are included in this summary.

\end{document}